\documentclass[twocolumn]{aastex63}

\newcommand{\chandra}{{\it Chandra}}
\newcommand{\xmm}{{\it XMM-Newton}}
\newcommand{\hst}{{\it HST}}
\newcommand{\kms}{km s$^{-1}$}
\newcommand{\Msun}{M$_{\odot}$}
\newcommand{\Rsun}{R$_{\odot}$}
\newcommand{\Lsun}{L$_{\odot}$}

\newcommand{\Lx}{L$_{\rm X}$}
\newcommand{\NH}{$N_{\rm H}$}
\newcommand{\flux}{erg s$^{-1}$ cm$^{-2}$}
\newcommand{\lum}{erg s$^{-1}$}
\newcommand{\vinf}{$v_{\infty}$}

\newcommand{\MBH}{M$_{\rm BH}$}
\newcommand{\Mstar}{M$_{\star}$}
\newcommand{\Tstar}{$T_{\star}$}
\newcommand{\Lstar}{$L_{\star}$}
\newcommand{\Rstar}{$R_{\star}$}
\newcommand{\massloss}{$\dot{M}_{\star}$}

\newcommand{\mrate}{M$_{\odot}$ yr$^{-1}$}

\usepackage{multirow}
\usepackage{xcolor}

\DeclareRobustCommand{\ion}[2]{%
\relax\ifmmode
\ifx\testbx\f@series
{\mathbf{#1\,\mathsc{#2}}}\else
{\mathrm{#1\,\mathsc{#2}}}\fi
\else\textup{#1\,{\mdseries\textsc{#2}}}%
\fi}

\newcommand{\HeIIuv}{\ion{He}{II} $\lambda$1640}
\newcommand{\HeIIop}{\ion{He}{II} $\lambda$4686}
\newcommand{\CIV}{\ion{C}{IV} $\lambda$1550}

\received{January 11, 2021}
\revised{February 9, 2021}
\accepted{February 13, 2021}
\submitjournal{ApJ}

\shorttitle{The Mass of NGC 300 X-1}
\shortauthors{Binder et al.}

\graphicspath{{./}{figures/}}

\begin{document}

\title{The Wolf-Rayet + Black Hole Binary NGC 300 X-1: What is the Mass of the Black Hole?}

\correspondingauthor{Breanna A. Binder}
\email{babinder@cpp.edu}

\author{Breanna A. Binder}
\affiliation{Department of Physics \& Astronomy, California State Polytechnic University, 3801 W. Temple Ave, Pomona, CA 91768, USA}

\author{Janelle M. Sy}
\affiliation{Department of Physics \& Astronomy, California State Polytechnic University, 3801 W. Temple Ave, Pomona, CA 91768, USA}

\author{Michael Eracleous}
\affiliation{Department of Astronomy \& Astrophysics and Institute for Gravitation and the Cosmos, The Pennsylvania State University, 525 Davey Lab, University Park, PA 16802, USA}

\author{Dimitris M. Christodoulou}
\affiliation{Lowell Center for Space Science and Technology, University of Massachusetts Lowell, 600 Suffolk Street, Lowell, MA 01854, USA}

\author{Sayantan Bhattacharya}
\affiliation{Lowell Center for Space Science and Technology, University of Massachusetts Lowell, 600 Suffolk Street, Lowell, MA 01854, USA}

\author{Rigel Cappallo}
\affiliation{Lowell Center for Space Science and Technology, University of Massachusetts Lowell, 600 Suffolk Street, Lowell, MA 01854, USA}

\author{Silas Laycock}
\affiliation{Lowell Center for Space Science and Technology, University of Massachusetts Lowell, 600 Suffolk Street, Lowell, MA 01854, USA}

\author{Paul P. Plucinsky}
\affiliation{Harvard-Smithsonian Center for Astrophysics, 60 Garden Street, Cambridge, MA 02138, USA}

\author{Benjamin F. Williams}
\affiliation{Department of Astronomy, University of Washington, Box 351580, Seattle, WA 98195, USA}

\begin{abstract}

We present new X-ray and UV observations of the Wolf-Rayet + black hole binary system NGC~300 X-1 with the \chandra\ X-ray Observatory and the {\it Hubble Space Telescope} Cosmic Origins Spectrograph. When combined with archival X-ray observations, our X-ray and UV observations sample the entire binary orbit, providing clues to the system geometry and interaction between the black hole accretion disk and the donor star wind. We measure a binary orbital period of 32.7921$\pm$0.0003 hr, in agreement with previous studies, and perform phase-resolved spectroscopy using the X-ray data. The X-ray light curve reveals a deep eclipse, consistent with inclination angles of $i=60-75^{\circ}$, and a pre-eclipse excess consistent with an accretion stream impacting the disk edge. We further measure radial velocity variations for several prominent FUV spectral lines, most notably \HeIIuv\ and \CIV. We find that the \ion{He}{II} emission lines systematically lag the expected Wolf-Rayet star orbital motion by a phase difference $\Delta \phi\sim0.3$, while \CIV\ matches the phase of the anticipated radial velocity curve of the Wolf-Rayet donor. We assume the \CIV\ emission line follows a sinusoidal radial velocity curve (semi-amplitude = 250 \kms) and infer a BH mass of 17$\pm$4 \Msun. Our observations are consistent with the presence of a wind-Roche lobe overflow accretion disk, where an accretion stream forms from gravitationally focused wind material and impacts the edge of the black hole accretion disk.

\end{abstract}


\section{Introduction} \label{sec:intro}
High mass X-ray binaries (HMXBs) are systems composed of a massive OB star and a neutron star (NS) or black hole (BH). Winds from the stellar companion are accreted onto the compact object, releasing copious amounts of X-ray radiation. A large fraction of the X-rays that are generated close to the compact object escape -- making these systems detectable as bright X-ray sources, with X-ray luminosities \Lx\ $\sim10^{36-39}$ \lum, but a significant fraction of X-rays will interact with material in the immediate environment and be reprocessed before becoming accessible to X-ray telescopes. X-ray reprocessing is ubiquitious in systems with accreting compact objects \citep{deJong+96,Suleimanov+03,Gierlinski+09}; in HMXBs, the wind of the donor star provides a reservoir of material in which X-ray reprocessing can occur. Observations of X-ray reprocessing in HMXBs have yielded important constraints on mass accretion mechanisms, the geometry of accreting systems, and the distribution and ionization state of the stellar wind material \citep{Aftab+19,Day+93}.

In this paper we take a closer look at the X-ray and UV properties of the Wolf-Rayet (WR) + BH HMXB NGC 300 X-1 (hereafter X-1). The high X-ray luminosity of the system \citep[$\sim5\times10^{38}$ \lum\ in the 0.35-8 keV energy band; ][]{Binder+15,Binder+11} and high fractional spectral variability rms value of $\sim$0.2 \citep{Earnshaw+17} are typical of accreting stellar mass BHs in a very high/steep power law state. Evidence of a soft thermal component in the X-ray spectrum has been interpreted as a possible signature of X-ray reprocessing by the photoionized stellar wind from the WR donor \citep{Carpano+07}.

Previous radial velocity (RV) measurements of the \HeIIop\ emission line indicated a BH mass (\MBH) of $\sim$ 20\Msun\ \citep{Crowther+10}, making X-1 host to the second-heaviest known stellar-mass black hole in an XRB \citep[after IC~10 X-1;][]{Laycock+15a,Laycock+15b,Prestwich+07,Silverman+08,Barnard+08}. However, subsequent \chandra\ \citep{Binder+15} and \xmm\ \citep{Carpano+19} observations revealed a phase shift between the X-ray eclipse light curve and the \HeIIop\ RV curve, suggesting that this emission line does not actually trace the motion of the Wolf-Rayet component and casting doubt on prior estimates of \MBH. A similar phase offset was first observed in Cyg~X-3 \citep{vanKerkwijk93} and interpreted as the result of strong X-ray irradiation of the stellar wind. A phase shift in the \HeIIop\ RV curve was similarly observed in IC~10 X-1 \citep{Laycock+15a,Laycock+15b}, further calling into question the reliability of BH mass estimates derived from this emission line. 

It has been suggested that the bright \HeIIop\ emission lines observed in these systems may instead be originating from the BH accretion disk or a region of the donor star's dense stellar wind that is shadowed from X-ray irradiation from the BH \citep{Binder+15,Laycock+15a,Laycock+15b,Carpano+19}. Additionally, \HeIIop\ emission has been observed originating from hot spots on compact object accretion disks, where an accretion stream flowing from the donor star through the L1 point impacts the outer edge of the accretion disk \citep[e.g., ][]{Marsh+90a,Marsh+90b}. Line-driven winds, such as those from WR stars and other hot stars \citep{Lucy+70,Castor+75}, are strongly affected by X-ray irradiation from a close accreting compact object \citep{Hatchett+77,Fransson+80}. At typical X-1 luminosities (\Lx\ of a few $\times10^{38}$ \lum), the WR wind is expected to be highly ionized \citep{Blondin+91,Manousakis+15}, and at very close orbital separations \citep[the orbital period of X-1 is $\sim$33 hr;][]{Carpano+07,Crowther+10,Carpano+19} may disrupt the wind acceleration mechanism to such an extent that a photoionization ``wake'' is formed \citep{Blondin+90,Blondin+94,Krticka+15}. This interaction between the X-ray irradiation from the BH and the stellar wind from the WR is expected to imprint itself on UV and X-ray spectra of HMXBs  \citep[as observed in some Galactic HMXBs, including Vela X-1 and 4U 1700-37,][]{Kaper+94,Schulz+02}.

We have obtained a new X-ray observation from the \chandra/ACIS-I instrument and four epochs of FUV spectroscopy with the {\it Hubble Space Telescope}'s Cosmic Origins Spectrograph (\hst/COS) to investigate the geometry of the X-1 binary, the mass-transfer and mass-accretion mechanisms operating in the system, and to attempt to place more reliable constraints on the mass of the BH. WR stars are bright UV sources, with strong emission from \HeIIuv\ and resonance lines that are not available in the optical spectra (such as \CIV) that can be directly compared to previous optical observations. However, even modest uncertainties in the binary orbital period \citep[$\sim$10 s; ][]{Carpano+18} will lead to large errors in determining the orbital phases at which both the archival optical spectra and new UV observations were obtained, translating into unacceptably large errors on any BH mass estimate. We therefore obtained a new X-ray observation with \chandra\ taken nearly simultaneously with the \hst\ observations. When combined with archival X-ray observations, our new observation extends the baseline of available X-ray observations to two decades. These new observations have allowed us to dramatically refine the orbital period measurement, calculate to high accuracy the orbital phases at which the UV and optical spectra were obtained, and estimate the mass of the X-1 BH.

The outline of this paper is as follows. In Section~\ref{sec:observations}, we describe the new and archival observations utilized in this work, along with details of our observing strategy and data processing. In Section~\ref{sec:period}, we refine the ephemeris of the X-1 binary, and in Section~\ref{section:Xray_spec} we analyze the X-ray spectra of X-1 as a function of binary orbital phase. We present an analysis of the FUV spectra as a function of orbital phase in Section~\ref{sec:UVanalysis}. In Section~\ref{sec:discussion} we derive new constraints on the X-1 binary parameters and present a physical model to interpret our observations (the updated BH mass estimate is discussed specifically in Section~\ref{section:BHmass}). We also discuss the assumptions and caveats to our results in that section. We conclude with a summary of our findings in Section~\ref{sec:summary}. Throughout this work, we assume a distance to NGC~300 of 2.0 Mpc \citep[derived from the tip of the red giant branch magnitudes observed by the \hst;][]{Dalcanton+09}. Assuming $H_0=70$ km s$^{-1}$ Mpc$^{-1}$, we estimate the galaxy's bulk recessional velocity is $\sim$141 \kms\ and the corresponding redshift is $z=0.00047$ \citep[which broadly agrees with direct measurements; e.g.,][]{Lauberts+89}.

\section{New and Archival Observations} \label{sec:observations}
NGC~300 has been previously observed with the \chandra\ Advanced CCD Imaging Spectrometer (ACIS-I) three times \citep{Binder+15,Binder+11} and with \xmm\ seven times, primarily for the purpose of studying the X-ray binary population of its host galaxy \citep{Binder+12}; two deep \xmm\ observations were performed in concert with {\it NuSTAR} specifically to study NGC~300 X-1, and serendipitously caught the neighboring pulsar-HMXB (the supernova ``impostor'' SN~2010da) in an ultraluminous state (after which the object became known as NGC~300 ULX-1; \citep{Walton+18,Carpano+18,Koliopanos+19}. To add to this archival data set, we have obtained new \chandra/ACIS-I observations of NGC~300 (observation ID 22375) and four coordinated observations with \hst/COS (program 15999) to perform FUV spectroscopy of X-1 throughout its orbital cycle. Table~\ref{tab:obs_log} summarizes the relevant information for all observations (new and archival) utilized in this work, which we have reprocessed as described below.

\begin{deluxetable*}{ccccccc}
\tablenum{1}
\tablecaption{Observation Log\label{tab:obs_log}}
\tablewidth{0pt}
\tablehead{
\colhead{} & \colhead{} & \colhead{Observation} & \colhead{Observation} &
\colhead{Exposure} & \colhead{New or} & Orbital  \\
\colhead{Observatory} & \colhead{Instrument} & \colhead{Date} & \colhead{ID \#} &
\colhead{Time} & \colhead{Archival?} & Phase$^a$ }
\decimalcolnumbers
\startdata
\xmm        & EPIC pn   & 27 Dec 2000   & 0112800201        & 34.6   & archival  & 0.60-0.93 \\
\xmm        & EPIC pn   & 01 Jan 2001   & 0112800101        & 27.5   & archival  & 0.46-0.63 \\
\xmm        & EPIC pn   & 22 May 2005   & 0305860401        & 29.4   & archival  & 0.80-1.08\\
\xmm        & EPIC pn   & 25 Nov 2005   & 0305860301        & 29.3   & archival  & 0.73-1.02 \\
\xmm        & EPIC pn   & 28 May 2010   & 0656780401        & 10.6   & archival  & 0.90-1.00 \\
\chandra    & ACIS-I    & 24 Sep 2010   & 12238             & 63.0 ks   & archival  & 0.64-1.18 \\
\chandra    & ACIS-I    & 16 May 2014   & 16028             & 64.2 ks   & archival  & 0.55-1.10 \\
\chandra    & ACIS-I    & 17 Nov 2014   & 16029             & 61.3 ks   & archival  & 0.76-1.29 \\
\xmm        & EPIC pn   & 17 Dec 2016   & 0791010101        & 110.3 ks  & archival  & 0.00-1.00 \\
\xmm        & EPIC pn   & 19 Dec 2016   & 0791010301        & 60.4 ks   & archival  & 0.06-0.73 \\
\chandra    & ACIS-I    & 26 Apr 2020   & 22375             & 47.4 ks   & {\bf new} & 0.76-1.17  \\
\hst        & COS/FUV   & 27 Apr 2020   & 15999-03          & 2 orbits  & {\bf new} & 0.738-0.761  \\
\hst        & COS/FUV   & 27 Apr 2020   & 15999-04          & 2 orbits  & {\bf new} & 0.981-1.004  \\
\hst        & COS/FUV   & 29 Apr 2020   & 15999-02          & 2 orbits  & {\bf new} & 0.485-0.507  \\
\hst        & COS/FUV   & 30 Apr 2020   & 15999-01          & 2 orbits  & {\bf new} & 0.261-0.283  \\
\enddata
\tablecomments{$^a$See Section~\ref{sec:period} for discussion of the X-1 ephemeris and orbital phase calculations.}
\end{deluxetable*}

\subsection{\chandra\ Data Reduction}
All \chandra\ data reduction was done in CIAO\footnote{\url{http://cxc.harvard.edu/ciao}} v4.12 \citep{Fruscione+06} and reprocessed from \texttt{evt1} using \texttt{chandra\_repro} and standard reduction procedures. Background light curves were extracted, but we found no evidence for strong background flares in any of the four observations. We created good time intervals (GTIs) using \texttt{lc\_clean}. All event data were filtered on these GTIs and restricted to the 0.5-7 keV energy range. The total usable \chandra\ exposure time was 235.9 ks, with the effective exposure times for the individual observations listed in Table~\ref{tab:obs_log}.

Exposure maps and exposure-corrected \chandra\ images were created using the \texttt{CIAO} task \texttt{fluximage}; we assumed spectral weights appropriate for a $\Gamma=1.7$ power law absorbed by the average foreground column density in the direction of NGC~300 \citep[\NH\ = 10$^{21}$ cm$^{-2}$; ][]{HI4PICollaboration+16}. All observations were corrected to the Solar System barycenter to account for differences in photon arrival times due to both \chandra's and the Earth's orbital motion using the task \texttt{axbary}. The point source detection tool \texttt{wavdetect} was used to detect X-ray sources in all four observations; as the source with the highest X-ray flux in NGC~300 at the times of the \chandra\ exposures, X-1 was robustly detected in all four observations.

\begin{figure}
    \centering
    \includegraphics[width=0.9\linewidth,clip,trim=1.2cm 6cm 1.2cm 6cm]{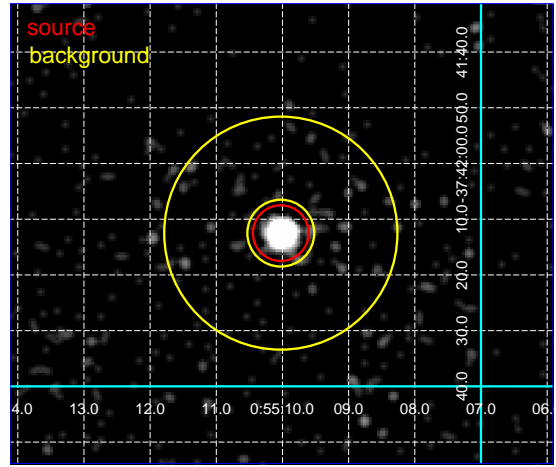}
    \caption{The X-ray image of X-1 from \chandra\ ObsID 22375. The red and yellow regions show the source and background regions, respectively, used to extract the \chandra\ light curves.}
    \label{fig:Xray_img}
\end{figure}

To extract light curves, we assumed a circular source region (5$^{\prime\prime}$ radius, chosen to enclose $\sim$90\% of the \chandra\ point spread function) centered on the \texttt{wavdetect} position of X-1. Circular annuli were used to define the background region, with an inner radius set to 6$^{\prime\prime}$. The outer radius was set such that the background region contained $\sim$100 counts; in all observations, this corresponds to an outer radius of $\sim$20$^{\prime\prime}$. No faint X-ray sources were detected within the background region by \texttt{wavdetect} in any of the four \chandra\ observations. Background-subtracted light curves were then extracted using \texttt{dmextract} and binned to 2 ks. An image of the 0.5-7 keV source, using the new \chandra\ observation ObsID 22375, is shown in Figure~\ref{fig:Xray_img}, with the source and background light curve extraction regions shown.

\subsection{\xmm\ Data Reduction}
All \xmm\ data reduction was performed with SAS 18.\footnote{\url{https://www.cosmos.esa.int/web/xmm-newton/sas}} The raw \texttt{evt1} PN data were reprocessed with \texttt{epproc} and  filtered on an energy range of 0.2-10 keV and PATTERN$<=$4. Background light curves were extracted using \texttt{evselect} and inspected for all observations, and good time intervals were generated to reject observation times exhibiting background flaring events. The total usable GTI for all \xmm\ observations was 302.1 ks, with the effective exposure times for the individual observations listed in Table~\ref{tab:obs_log}.

We used the \chandra\ position to define the source and background extraction regions in the \xmm\ field. A circular region of radius 20$^{\prime\prime}$ was used for the source region. Due to the brightness of nearby ($\sim$1$^{\prime}$) NGC~300 ULX-1 in two of the observations, we defined a circular background region positioned near X-1 on the same PN chip that was free from stray light contamination from ULX-1. We used a light curve time resolution of 100 s. The SAS tool \texttt{epiclcorr} was used to correct the light curves for bad pixels, PSF variation at the location of X-1, as well as dead time, and vignetting. 

\subsection{\hst/COS FUV Spectroscopy}
We successfully obtained eight orbits with \hst/COS to observe NGC~300 X-1 at four different phases of the binary's orbit. Two observations were obtained $\sim$8 hours apart on 2020 April 27, and the remaining two observations were obtained on 2020 April 29 and 30.  We used the G140L grating with a central wavelength of 1280 \AA\ in \texttt{TIME-TAG} mode. This configuration produces two spectral segments, covering $\sim$915--2150 \AA\ with a gap at 1193--1266 \AA; our spectra therefore do not cover Ly$\alpha$. The spectral resolution is $\sim$0.08 \AA, and we estimate errors using Poisson statistics on the gross counts within each bin using the Gehrels approximation \citep{Gehrels86}.

\begin{figure}
    \centering
    \includegraphics[width=1\linewidth]{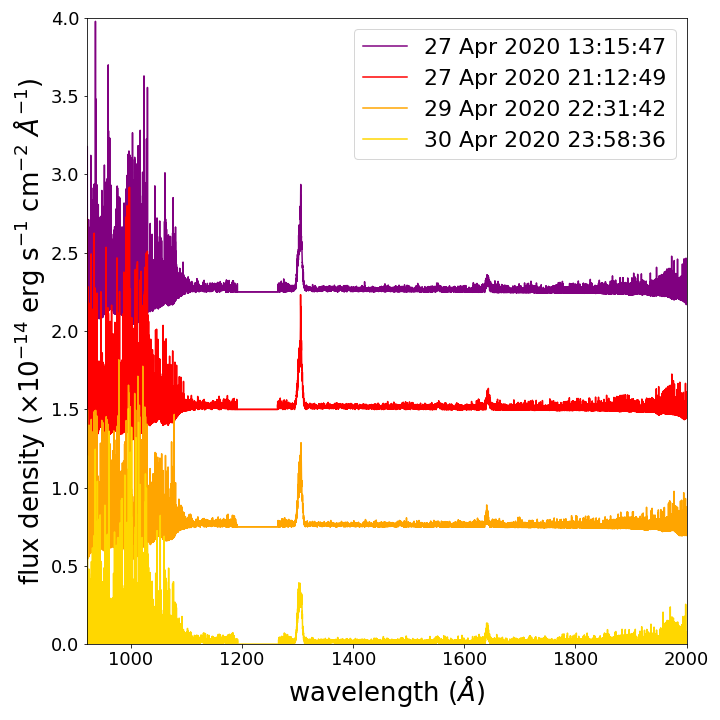}
    \caption{The \hst/COS FUV spectra of NGC~300 X-1, color-coded by observation date and start time. Observations have been given an arbitrary flux offset for clarity.}
    \label{fig:COS_spectra}
\end{figure}

The spectra of all four observations are shown in Figure~\ref{fig:COS_spectra}. The spectral signal deteriorates towards the red and blue edges of the spectrum. Bright geocoronal airglow lines of \ion{O}{I} at 1302 \AA\ and 1305 \AA\ are clearly seen, along with several fainter emission features intrinsic to the X-1 system. We discuss the FUV spectrum further in Section~\ref{sec:UVanalysis}.

\section{Refining the Orbital Period} \label{sec:period}
\begin{figure}
    \centering
    \includegraphics[width=1\linewidth]{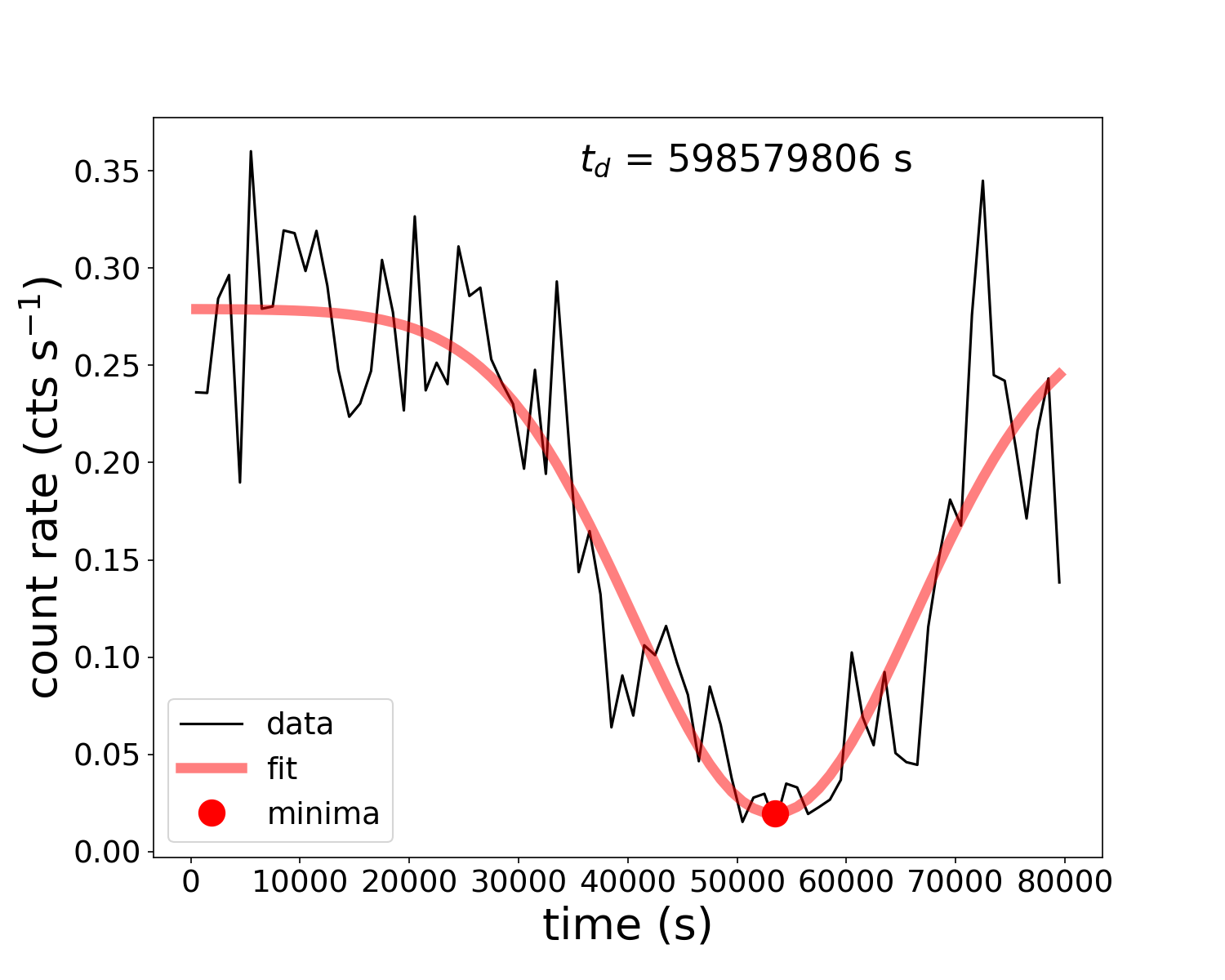}
    \caption{Finding the deepest part of the X-ray dip with \xmm\ observation from 19 December 2016 (time 0 indicates the beginning of the observation). We defined the dip minimum as corresponding to phase $\phi=0.5$.}
    \label{fig:find_dip_min}
\end{figure}

In order to refine the orbital period measurement of the X-1 binary, we first aimed to phase-fold all four \chandra\ observations with the deep \xmm\ observations. The \xmm\ observation obtained on 19 December 2016 captured the entire extent of the X-ray dip; we therefore used the light curve for this observation to identify, to the highest precision possible, the exact minimum of the dip, which we will define as the phase $\phi=0.5$. As shown in Figure~\ref{fig:find_dip_min}, an inverted Gaussian superimposed on a constant continuum count rate provided a good fit to the observed light curve, with the deepest part of the X-ray dip occurring at $t_{\rm d}$=598\,579\,806 s (mission seconds, corresponding to MJD 57742.00701).

\begin{figure*}[!ht]
    \centering
    \includegraphics[width=1\linewidth]{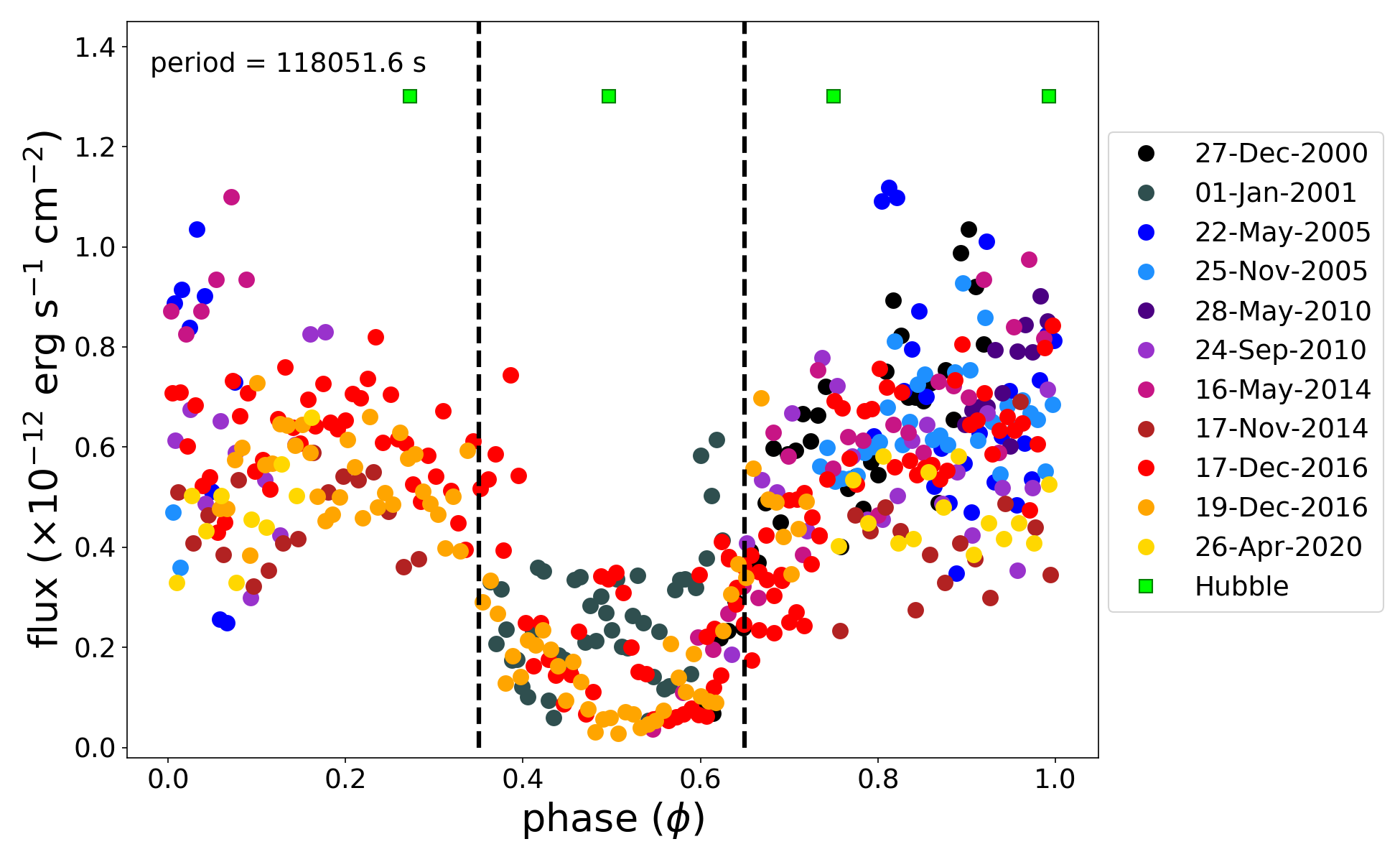}
    \caption{All \chandra\ and \xmm\ light curves folded on the best orbital period, 32.7921$\pm$0.0003 hr. The green squares indicate the phases at which the \hst\ observations were obtained; the width of the squares indicates the spread in orbital phase due to the \hst\ exposure time interval. The vertical dashed lines show $\phi=0.35$ and 0.65, indicating the approximate extent of the X-ray eclipse.}
    \label{fig:best_period}
\end{figure*}

The combined \chandra\ and \xmm\ observations span a $\sim$20 year period, with only $\sim$2 days separating the two deepest \xmm\ observations. Given the long baseline of observations available for the system, maintaining phase connectivity would require knowing the orbital period to extremely high precision. The orbital period of X-1 was first determined to be $\sim$32.8 hr from X-ray monitoring from the {\it Neil Gehrels Swift Observatory} \citep{Carpano+07}, and optical RV measurements of the \HeIIop\ emission line \citep{Crowther+10} were consistent with the X-ray orbital period. We therefore explored a set of trial periods ranging from 32.0--34.0 hours. The arrival times $T$ of each \chandra\ and \xmm\ event were transformed to orbital phase $\phi$ using the 19 December 2016 mid-dip time to define $\phi=0.5$ (to avoid negative phases, we subtracted 100\,000 cycles of the trial period, as was done in \citet{Laycock+15a}:

\begin{equation}
    \phi = \frac{\left(T-t_d-100\,000 P\right)}{P}
\end{equation}

The time systems used by both \chandra\ and \xmm\ share the same reference point (time elapsed since 1998.0 TT), so no additional time conversion between light curves is required between missions. We converted the count rates measured in the \chandra\ and \xmm\ light curves to an (absorbed) 0.5-7 keV flux, assuming a basic power law spectrum with $\Gamma=1.7$ subject to an absorbing column of \NH\ $=10^{21}$ cm$^{-2}$. A more detailed analysis of the evolution of the X-ray luminosity over the orbital period is discussed in Section~\ref{section:Xray_spec}. Folded light curves were initially generated in 1 s intervals and visually inspected. A period range of 118049 s--118055 s yielded relatively good phase connectivity; we therefore generated a new set of folded light curves within this range using 0.2 s increments and created a movie of the resulting phase-folded light curve (available in the online version of the paper). 

The best period is 32.7921 hr, shown in Figure~\ref{fig:best_period}, in excellent agreement with previous studies \citep{Carpano+19}. We use our solution to determine the orbital phases at which the \hst\ observations were taken; these are shown as green squares on Figure~\ref{fig:best_period}. A change in the orbital period of $\pm$1.2 s causes the dip structure between $\phi=0.35-0.65$ to lose coherence; we therefore adopt 1.2 s (0.0003 hr) as the uncertainty in the orbital period. 

\section{Phase-Resolved X-ray Spectroscopy}\label{section:Xray_spec}
As shown in Figure~\ref{fig:best_period}, the \hst/COS observations were obtained at $\phi=0.273$, 0.497, 0.750, and 0.993, and multiple \chandra\ and \xmm\ observations are available across the entire binary orbit. We therefore extracted four sets of X-ray spectra that correspond to the \hst\ phases. We define these four phases as $\phi=0.00$ (spanning phases 0.875--0.125), 0.25 (spanning phases 0.125--0.375, corresponding to the X-ray dip ingress), 0.50 (corresponding to the X-ray dip at 0.375--0.625), and 0.75 (corresponding to the X-ray dip egress, spanning 0.625--0.875). All \chandra\ spectra were extracted using the CIAO task \texttt{specextract} in the energy range of 0.5-7 keV and binned to contain at least ten counts per bin. Phase-resolved \xmm\ PN spectra were extracted in the energy range 0.3-10 keV using \texttt{evselect} and binned to contain at least 25 counts per bin.

We simultaneously fitted all available X-ray spectra for in each phase bin with XPSEC v.12.9.1 \citep{Arnaud96}. All reported uncertainties correspond to the 90\% confidence interval. Our spectral model consists of two thermal plasma components (\texttt{mekal}) to model X-ray emission originating in the wind of the WR donor, a Comptonized continuum component with thermal seed photons (\texttt{comptt}) to represent the BH accretion disk, an absorbing column due to the Galaxy \citep[\texttt{tbabs}, fixed at $10^{21}$ cm$^{-2}$;][]{HI4PICollaboration+16}, and a partial covering absorbing column (\texttt{pcfabs}) to account for additional absorption of soft X-rays as the BH moves through the WR wind. The final XSPEC model is therefore \texttt{tbabs*pcfabs*(mekal+mekal+comptt)}, and the free parameters of the fit are the partial covering absorbing column density \NH, the partial covering fraction $f_{\rm cov}$, the temperatures of the two thermal plasmas ($kT_1$ and $kT_2$), the ``seed'' temperature $T_0$ of the soft photons that interact with the Comptonizing plasma (which has a temperature of $kT$), and the optical depth $\tau$ of the Comptonizing plasma.

\begin{figure*}
\centering
\begin{tabular}{cc}
    \includegraphics[width=0.49\linewidth,clip=true,trim=0.5cm 0.75cm 1cm 1.5cm]{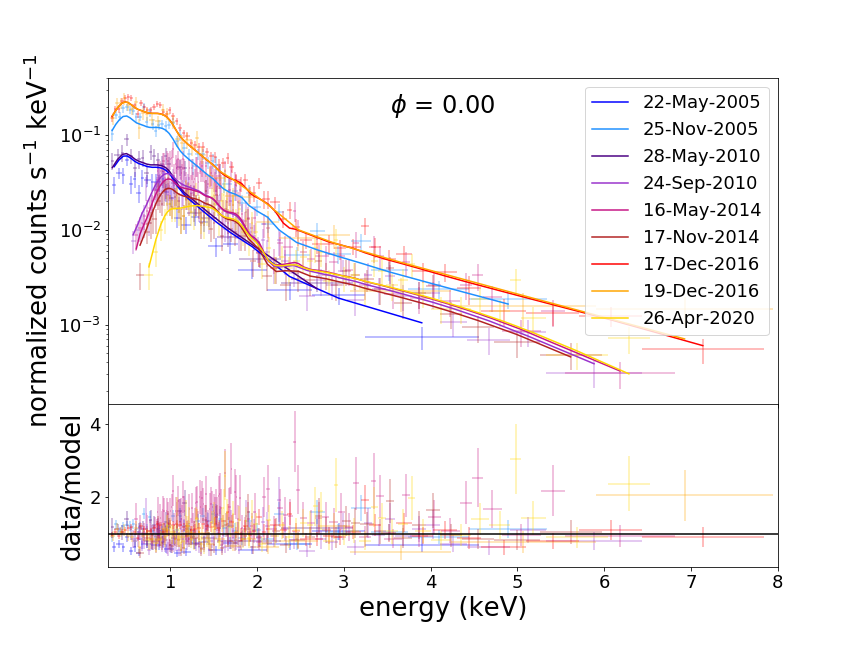} &  
    \includegraphics[width=0.49\linewidth,clip=true,trim=0.5cm 0.75cm 1cm 1.5cm]{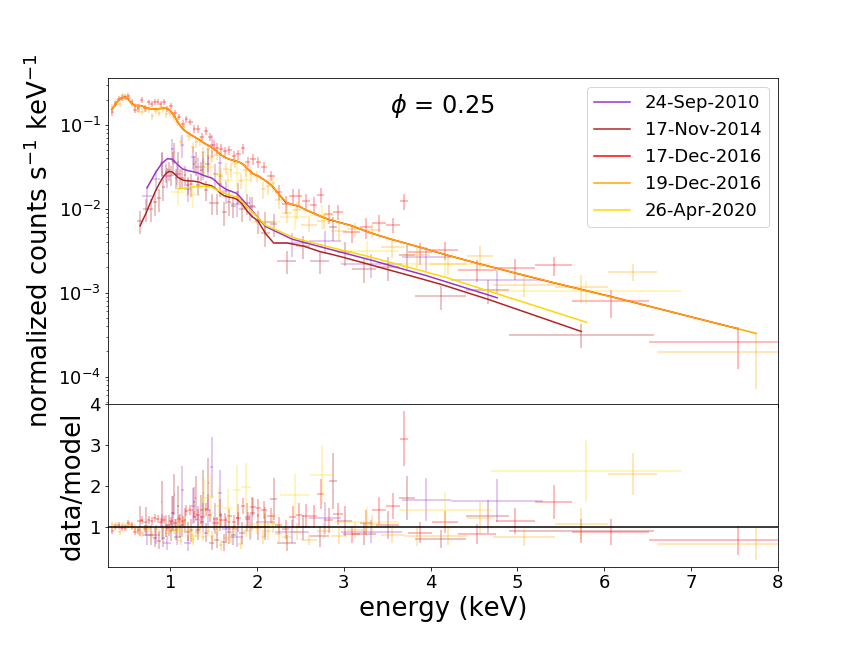} \\
    \includegraphics[width=0.49\linewidth,clip=true,trim=0.5cm 0.75cm 1cm 1.5cm]{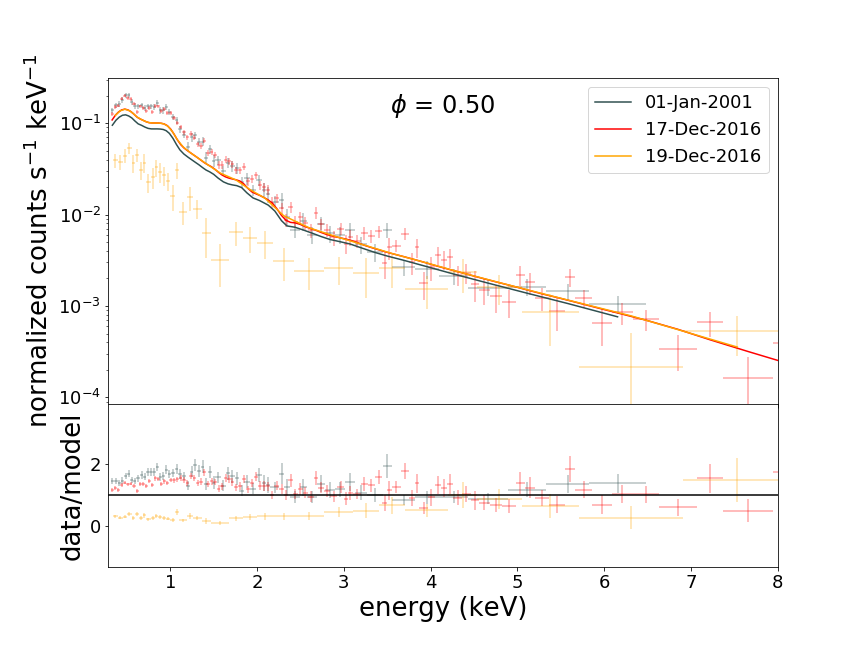} &  
    \includegraphics[width=0.49\linewidth,clip=true,trim=0.5cm 0.75cm 1cm 1.5cm]{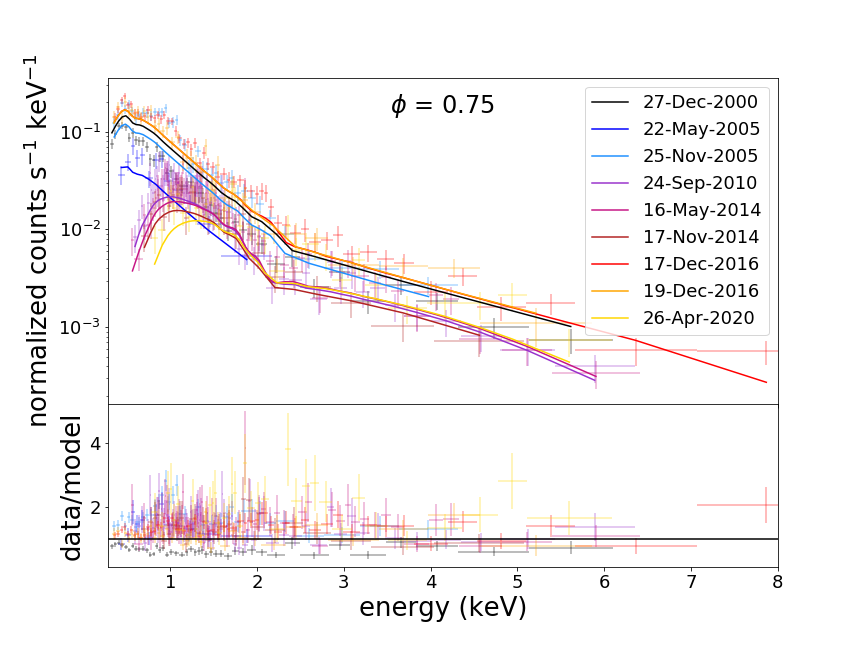} \\
\end{tabular}
    \caption{Phase-resolved X-ray spectroscopy with \chandra\ and \xmm. All available spectra are fit simultaneously, with all free parameters tied between observations. Colors are the same as in Figure~\ref{fig:best_period}.}
    \label{fig:phase_res_Xspectra}
\end{figure*}

Initially, all free parameters in our model were free to vary but they were tied to the same value for all of the simultaneously fitted spectra. Figure~\ref{fig:phase_res_Xspectra} shows the results of our simultaneous fitting as a function of orbital phase. Although the quality of the simultaneous fits was generally good ($\chi^2$/dof $<$ 2) for observations at $\phi=0.00$ and 0.25, keeping parameters tied between observations yielded poor fits to the data at $\phi=0.50$ and 0.75. The best-fit parameters are summarized in Table~\ref{tab:phase_res_Xspectra}. The quality of the spectral fits was significantly driven by the best-fit values of some parameters (such as \NH\ and the partial covering fraction $f_{\rm cov}$, and possibly the optical depth of the Comptonizing plasma $\tau$), while being insensitive to the values of other parameters.

\begin{deluxetable*}{ccccc}
\tablenum{2}
\tablecaption{Phase-Resolved X-ray Spectroscopy Best-Fit Parameters\label{tab:phase_res_Xspectra}}
\tablehead{
                    & \multicolumn{4}{c}{Orbital Phase$^a$} \\ \cline{2-5}
\colhead{Parameter} & \colhead{0.00} & \colhead{0.25} & \colhead{0.50} & \colhead{0.75}  
}
\decimalcolnumbers
\startdata
\NH\ ($10^{22}$ cm$^{-2}$)   & 4.1$^{+1.6}_{-1.1}$      & 2.1$^{+0.7}_{-0.6}$   & 3.1$^{+0.8}_{-0.6}$      & 5.1$^{+2.2}_{-1.4}$ \\
$f_{\rm cov}$               & 0.56$^{+0.05}_{-0.06}$    & 0.57$\pm$0.05         & 0.70$^{+0.08}_{-0.12}$    & 0.66$^{+0.08}_{-0.11}$ \\
$kT_{\rm 1}$ (keV)          & $<$0.2                    & \nodata               & $<$0.18                   & $<$0.25 \\
$kT_{\rm 2}$ (keV)          & 0.8$\pm$0.1               & 0.9$\pm$0.1           & 0.86$^{+0.20}_{-0.11}$    & 0.82$^{+0.51}_{-0.29}$ \\
soft photon $T_0$ (keV)     & $<$0.08                   & $<$0.07               & $<$0.09                   & $<$0.08 \\
plasma $kT$ (keV)           & 46$\pm$19                 & 51$\pm$16             & 57$\pm$13                 & 50$\pm$12 \\
optical depth $\tau$        & 0.19$\pm$0.02             & 0.14$\pm$0.01         & 0.12$^{+0.02}_{-0.01}$    & 0.14$\pm$0.01 \\
$\chi^2$/dof                & 987.8/493                 & 353.9/200             & 4254.8/436                & 1873.5/436 \\
\enddata
\tablecomments{$^a$The range of phases in each bin are described in Section~\ref{section:Xray_spec} of the text. Uncertainties correspond to the 90\% confidence interval.}
\end{deluxetable*}

We therefore re-fitted the X-ray spectra at each phase with the following parameters frozen: the two thermal plasma temperatures are fixed to $kT_1=0.1$ kev and $kT_2=0.8$ keV and the soft photon and Comptonizing plasma temperatures are set to $T_0=0.1$ keV and $kT=50$ keV, respectively. In this round of spectral fitting, we allow the partial covering absorbing column, covering fraction, and optical depth of the Comptonizing plasma (\NH, $f_{\rm cov}$, and $\tau$ respectively) to vary between observations within each phase bin. In addition, for each observation we measure the unabsorbed 0.3-10 keV luminosity (\Lx), as well as the fluxes originating in the two thermal plasmas (which we refer to as the ``thermal'' component, likely associated with the WR star winds, as we discuss below) and the Comptonized component (which we call the ``Compton'' component, likely associated with the BH accretion disk). 

\begin{figure*}
\centering
\begin{tabular}{cc}
    \includegraphics[width=0.49\linewidth,clip=true,trim=0.5cm 0.75cm 1cm 1.5cm]{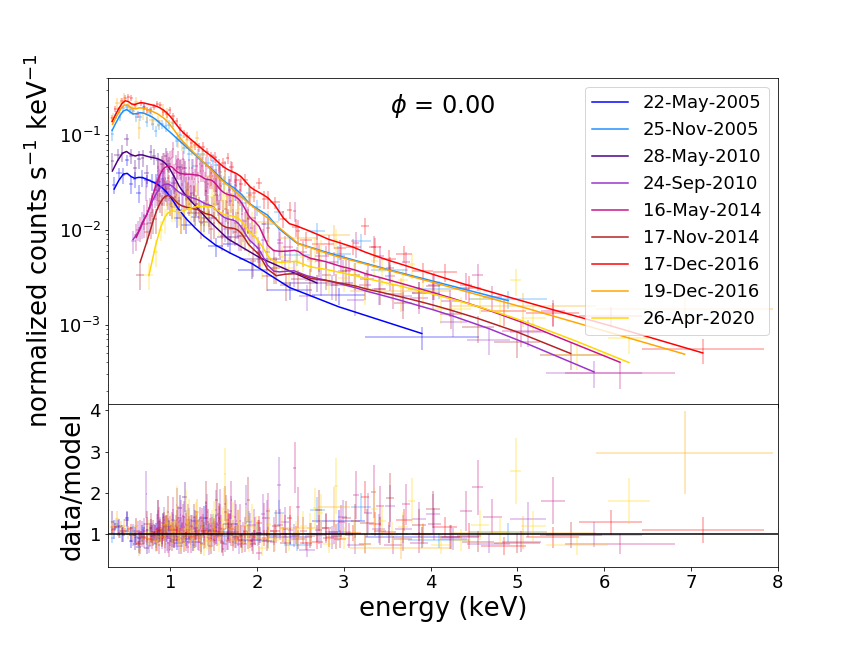} & 
    \includegraphics[width=0.49\linewidth,clip=true,trim=0.5cm 0.75cm 1cm 1.5cm]{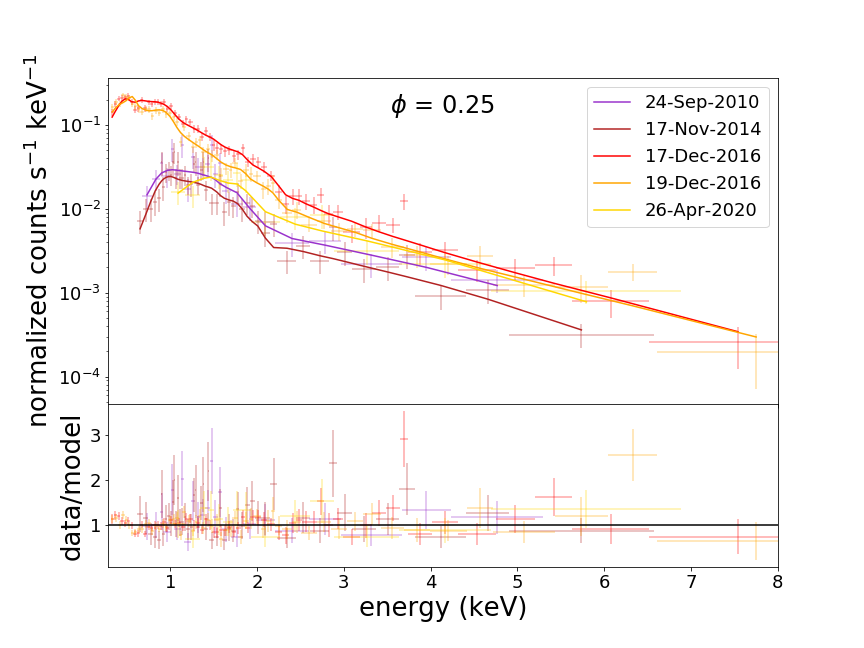} \\
    \includegraphics[width=0.49\linewidth,clip=true,trim=0.5cm 0.75cm 1cm 1.5cm]{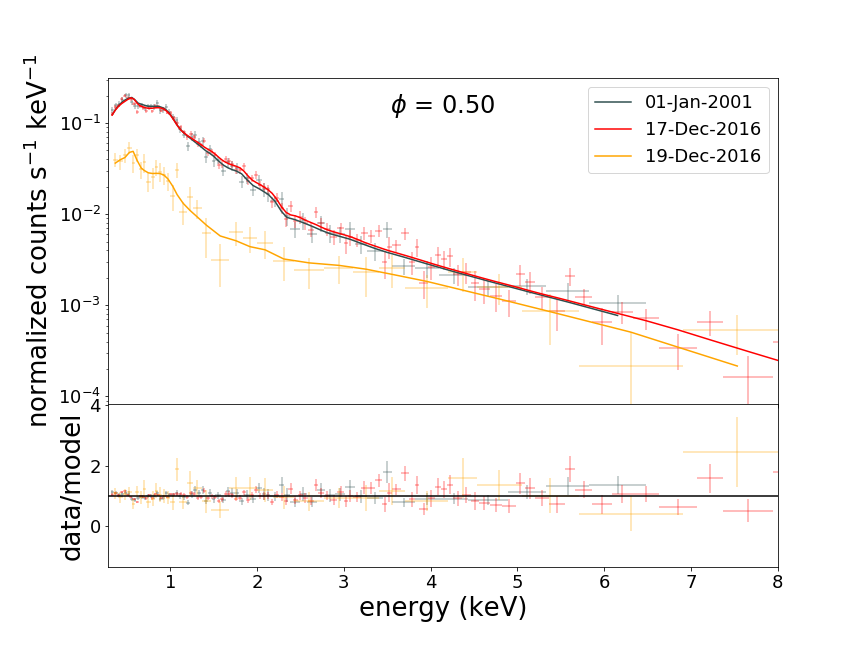} &  
    \includegraphics[width=0.49\linewidth,clip=true,trim=0.5cm 0.75cm 1cm 1.5cm]{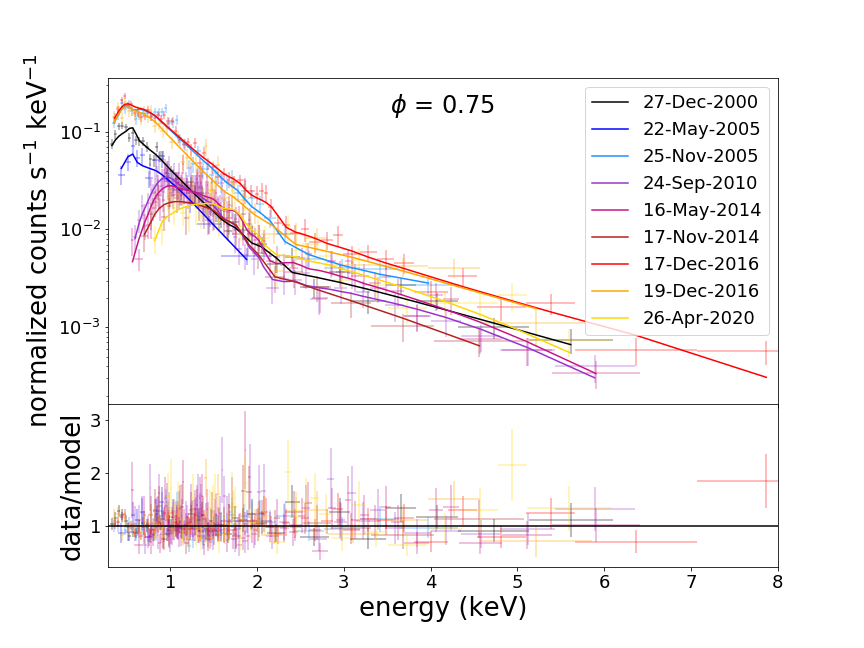} \\
\end{tabular}
    \caption{Phase resolved X-ray spectroscopy with \chandra\ and \xmm. All spectra were fit individually with the same spectral model, but the partial covering fraction $f_{\rm cov}$ and \NH and the Comptonizing plasma optical depth $\tau$ were allowed to vary between observations. All other fit parameters were fixed (see text for details). Colors are the same as in Figure~\ref{fig:best_period}.}
    \label{fig:phase_res_Xspectra_single}
\end{figure*}

Figure~\ref{fig:phase_res_Xspectra_single} shows the results of our second round of spectral fitting, and the best-fit parameters for individual observations as a function of orbital phase are summarized in Table~\ref{tab:phase_res_Xspectra_single}. Some observations, when resolved into specific phase bins, did not contain a sufficient number of counts to directly constrain one or more of the free fit parameters; in these cases, we report upper or lower limits on the fit parameters as appropriate. We additionally report the average fit parameters (omitting observations with only limits on the fit parameters), as well as the percentage of the X-ray luminosity originating in the thermal component.

\begin{deluxetable*}{cccccccc}
\tablenum{3}
\tablecaption{Individual Observation Phase-Resolved X-ray Spectroscopy Best-Fit Parameters\label{tab:phase_res_Xspectra_single}}
\tablewidth{0pt}
\tablehead{
\colhead{Observation} & \colhead{\NH} &  &  & \colhead{log$F_{\rm 0.3-10}^{\rm thermal}$} & \colhead{log$F_{\rm 0.3-10}^{\rm Compton}$} & \colhead{\Lx} (\% thermal) & \\
\colhead{Date}        & \colhead{($10^{22}$ cm$^{-2}$)} & \colhead{$f_{\rm cov}$} & \colhead{$\tau$} & ([\flux]) & ([\flux]) & ($\times10^{38}$ \lum) & \colhead{$\chi^2$/dof}  
}
\decimalcolnumbers
\startdata
\multicolumn{8}{c}{phase 0.00} \\
\hline
22-May-05 & 4.3$^{+2.1}_{-1.1}$ & 0.87$\pm$0.02 & 0.06$\pm$0.02 & -12.81$^{+0.50}_{-0.20}$ & -11.54$\pm$0.05 & $14.6^{+0.6}_{-0.2}$ (5.1) & 20.8/23 \\
25-Nov-05 & 7.5$^{+3.9}_{-2.9}$ & 0.60$^{+0.11}_{-0.20}$ & 0.12$\pm$0.02 & -14.30$^{+1.34}_{-0.74}$ & -11.70$^{+0.12}_{-0.17}$ & $9.8^{+0.9}_{-0.5}$ (0.2) & 56.2/37 \\
28-May-10 & 6.9$^{+4.5}_{-1.8}$ & 0.91$\pm$0.02 & 0.06$^{+0.03}_{-0.02}$ & -12.16$^{+0.38}_{-0.54}$ & -11.17$\pm$0.05 & $35.9^{+1.1}_{-1.6}$ (9.2) & 27.2/15 \\
24-Sep-10 & 2.9$^{+3.2}_{-1.1}$ & 0.57$^{+0.11}_{-0.10}$ & 0.16$^{+0.08}_{-0.05}$ & -12.41$^{+0.38}_{-1.08}$ & -12.00$^{+2.70}_{-0.06}$ & $6.7^{+0.2}_{-0.1}$ (27.8) & 60.1/72 \\
16-May-14 & $<$0.34 & $>$0.39 & 0.26$^{+0.06}_{-0.05}$ & -11.97$^{+0.96}_{-1.78}$ & -12.07$^{+0.11}_{-0.09}$ & $9.5^{+0.8}_{-1.4}$ (54.0) & 98.4/96 \\
17-Nov-14 & 3.7$^{+2.5}_{-1.3}$ & 0.63$^{+0.20}_{-0.30}$ & 0.20$^{+0.30}_{-0.12}$ & -12.75$^{+0.16}_{-0.26}$ & -11.97$^{+0.40}_{-0.31}$ & $6.0\pm0.2$ (14.2) & 43.7/55 \\
17-Dec-16 & 3.1$^{+5.2}_{-1.4}$ & 0.53$^{+0.18}_{-0.26}$ & 0.12$\pm$0.04 & -13.11$^{+0.42}_{-0.71}$ & -11.83$^{+0.17}_{-0.16}$ & $7.6^{+0.3}_{-0.4}$ (4.9) & 71.2/55 \\
19-Dec-16 & 6.8$^{+4.1}_{-3.3}$ & 0.71$\pm$0.02 & 0.09$\pm$0.03 & -13.03$\pm$0.43 & -11.69$^{+0.06}_{-0.07}$ & $10.3\pm0.3$ (4.3) & 37.9/23 \\
26-Apr-20 & 1.6$^{+5.0}_{-1.4}$ & 0.29$^{+0.36}_{-0.29}$ & 0.35$^{+0.32}_{-0.19}$ & -12.95$^{+0.19}_{-0.58}$ & -12.18$^{+0.27}_{-0.14}$ & $3.8^{+0.1}_{-0.2}$ (14.1) & 60.4/61 \\
{\bf Average}   & 4.0$^{+4.0}_{-1.7}$   & 0.59$^{+0.15}_{-0.18}$    & 0.15$^{+0.11}_{-0.07}$ & -12.96$^{+0.42}_{-0.60}$ & -11.85$^{+0.57}_{-0.14}$ & 8.6$^{+0.5}_{-0.4}$ (10.0) & \\
\hline
\multicolumn{8}{c}{phase 0.25} \\
\hline
24-Sep-10 & 32.2$\pm$13.9 & $<$0.82 & 0.31$^{+0.35}_{-0.12}$ & $<$-12.75 & -11.78$^{+0.41}_{-0.47}$ & $8.1\pm0.3$ ($<$10.5) & 21.5/16  \\
17-Nov-14 & 3.2$^{+3.9}_{-3.0}$ & $<$0.74 & 0.19$^{+0.24}_{-0.14}$ & -13.21$^{+0.41}_{-1.50}$ & -12.07$^{+0.52}_{-0.17}$ & $4.4^{+0.2}_{-0.5}$ (6.7) & 51.5/42 \\
17-Dec-16 & 1.4$^{+0.6}_{-0.5}$ & 0.65$^{+0.18}_{-0.23}$ & 0.09$^{+0.06}_{-0.04}$ & -12.85$\pm$0.50 & -11.73$^{+0.20}_{-0.21}$ & $10.0\pm0.4$ (6.8) & 79.7/55 \\
19-Dec-16 & 3.1$^{+1.8}_{-0.9}$ & 0.76$^{+0.11}_{-0.18}$ & 0.07$^{+0.04}_{-0.03}$ & -12.98$^{+0.36}_{-0.64}$ & -11.66$^{+0.22}_{-0.21}$ & $11.2^{+0.4}_{-0.6}$ (4.5) & 99.7/49 \\
26-Apr-20 & $<$3.3 & $<$0.86 & 0.19$^{+0.14}_{-0.10}$ & -12.24$^{+0.98}_{-4.07}$ & -11.89$^{+1.29}_{-0.98}$ & $9.2^{+1.2}_{-3.2}$ (29.9) & 9.5/6 \\
{\bf Average}   & 2.5$^{+1.5}_{-1.6}$ & 0.73$^{+0.09}_{-0.44}$ & 0.13$^{+0.11}_{-0.07}$ & -12.76$^{+0.56}_{-1.32}$ & -11.75$^{+0.44}_{-0.38}$ & 10.0$^{+0.6}_{-1.1}$ (12.0) & \\
\hline
\multicolumn{8}{c}{phase 0.50} \\
\hline
01-Jan-01   & 2.9$^{+2.5}_{-1.0}$ & 0.44$\pm$0.07 & 0.16$\pm$0.02 & -12.57$^{+0.22}_{-0.30}$ & -11.99$\pm$0.02 & $6.2\pm0.1$ (20.8) & 67.7/54 \\
17-Dec-16   & 1.7$^{+1.0}_{-0.6}$ & 0.49$^{+0.15}_{-0.25}$    & 0.17$^{+0.06}_{-0.05}$ & -12.52$^{+0.19}_{-0.20}$ & -12.05$\pm$0.13 & $5.8\pm0.1$ (24.9) & 173.2/91 \\
19-Dec-16   & 6.4$^{+3.0}_{-1.5}$ & 0.94$\pm$0.02 & 0.10$^{+0.10}_{-0.04}$ & -12.08$^{+0.19}_{-0.35}$ & -11.91$^{+0.08}_{-0.06}$ & $10.1^{+0.2}_{-0.3}$ (39.4) & 24.2/27 \\
{\bf Average}   & 3.3$^{+1.8}_{-1.9}$ & 0.62$^{+0.10}_{-0.15}$ & 0.15$^{+0.07}_{-0.04}$ & -12.39$^{+0.19}_{-0.26}$ & -12.00$^{+0.10}_{-0.09}$ & 7.2$^{+0.1}_{-0.2}$ (28.4) \\
\hline
\multicolumn{8}{c}{phase 0.75} \\
\hline
27-Dec-00   & 6.1$^{+6.8}_{-2.4}$ & 0.79$^{+0.10}_{-0.20}$ & 0.10$^{+0.04}_{-0.03}$ & -12.39$^{+0.13}_{-0.19}$ & -11.91$^{+0.30}_{-0.29}$ & $7.9\pm0.2$ (24.7) & 47.3/25 \\
22-May-05   & 0.2$\pm$0.1 & $>$0.73 & $<$0.09 & -12.09$^{+0.16}_{-0.26}$ & -11.98$^{+0.20}_{-0.21}$ & $16.7^{+0.4}_{-0.5}$ (23.3) & 9.5/5 \\
25-Nov-05   & $>$8.5 & 0.71$^{+0.26}_{-0.23}$ & 0.12$^{+0.03}_{-0.02}$ & $<$-12.29 & -11.57$^{+0.76}_{-0.48}$ & $14.8^{+0.1}_{-1.2}$ ($<$16.6) & 57.7/33 \\
24-Sep-10   & 7.1$^{+5.4}_{-2.4}$ & 0.69$^{+0.16}_{-0.21}$ & 0.11$^{+0.12}_{-0.05}$ & -12.97$^{+0.25}_{-0.65}$ & -11.77$^{+0.35}_{-0.40}$ & $8.8^{+0.3}_{-0.5}$ (5.8) & 56.2/68 \\
16-May-14   & 2.0$^{+2.0}_{-1.6}$ & 0.33$^{+0.22}_{-0.09}$ & 0.21$^{+0.16}_{-0.05}$ & -12.68$^{+0.77}_{-0.57}$ & -12.12$^{+0.39}_{-0.17}$ & $4.7^{+0.3}_{-0.2}$ (21.3) & 61.4/71 \\
17-Nov-14   & 2.2$^{+2.9}_{-0.9}$ & 0.74$^{+0.09}_{-0.15}$ & 0.06$^{+0.04}_{-0.02}$ & $<$-12.46 & -11.68$^{+0.12}_{-0.16}$ & $10.0^{+0.1}_{-0.3}$ ($<$16.6) & 15.2/22 \\
17-Dec-16   & 3.5$^{+3.0}_{-1.3}$ & 0.58$^{+0.16}_{-0.30}$ & 0.13$^{+0.06}_{-0.04}$ & $<$-12.70 & -11.87$^{+0.20}_{-0.22}$ & $7.0^{+0.1}_{-0.3}$ ($<$13.6) & 64.1/48 \\
19-Dec-16   & 6.1$^{+7.9}_{-2.7}$ & 0.81$^{+0.05}_{-0.09}$ & 0.08$^{+0.04}_{-0.03}$ & $<$-12.30 & -11.59$^{+0.12}_{-0.16}$ & $13.2^{+0.1}_{-0.4}$ ($<$18.2) & 36.5/29 \\
26-Apr-20   & $<$5.1 & 0.73$\pm$0.08 & 0.10$^{+0.41}_{-0.04}$ & -11.04$^{+0.70}_{-2.44}$ & -11.66$^{+0.12}_{-0.59}$ & $54.8^{+0.35}_{-1.24}$ (79.7) & 36.5/31 \\
{\bf Average}   & 4.6$^{+4.7}_{-2.5}$ & 0.65$^{+0.14}_{-0.18}$ & 0.12$^{+0.11}_{-0.04}$ & $<$-12.43 & -11.81$^{+0.28}_{-0.25}$ & 8.9$\pm$0.2 ($<$16.7) \\
\enddata
\tablecomments{\Lx\ refers to the unabsorbed, 0.3-10 keV luminosity derived from the best-fit model.}
\end{deluxetable*}

\begin{figure*}
    \centering
    \includegraphics[width=0.85\linewidth,clip=true,trim=1cm 1.5cm 1cm 3cm]{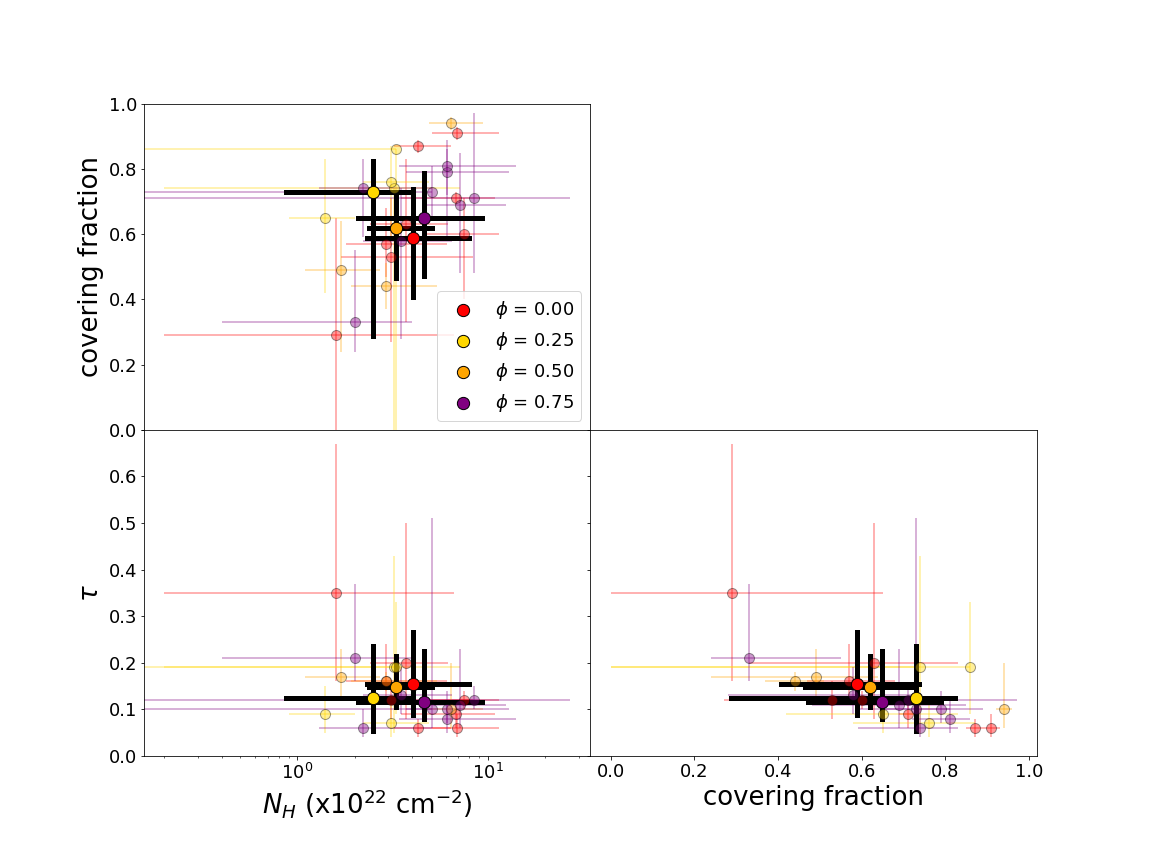}
    \caption{Comparison of \NH, partial covering fraction, and the Comptonizing plasma optical depth $\tau$. Individual X-ray observations are color-coded according to the corresponding orbital phases spanned by the \hst\ observations. The average values are shown with thick, black error bars.}
    \label{fig:parameter_comparison}
\end{figure*}

\begin{figure}
    \centering
    \includegraphics[width=1\linewidth,clip=true,trim=1cm 1.25cm 1cm 2cm]{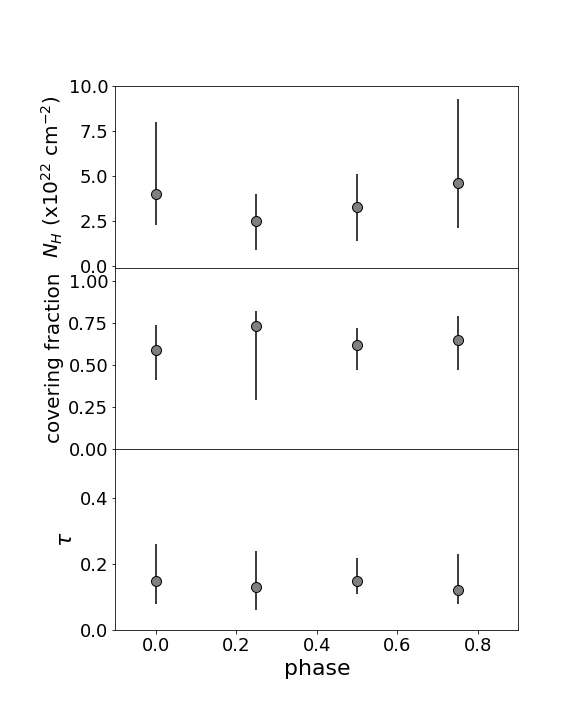}
    \caption{Average fit parameter values $N_{\rm H}$ (top), partial covering fraction (middle), and the optical depth $\tau$ (bottom) as a function of phase.}
    \label{fig:parameters_by_phase}
\end{figure}

Figure~\ref{fig:parameter_comparison} compares each of the three free fit parameters against the other two; we see no strong evidence for correlation between fit parameters. The average values of \NH, $f_{\rm cov}$, and $\tau$ are largely consistent across orbital phase (see Figure~\ref{fig:parameters_by_phase}). These results imply that bulk structure of the WR wind and BH accretion disk (or their orientation to our line-of-sight) are not changing significantly over the binary orbital phase.

\begin{figure}
    \centering
    \includegraphics[width=1\linewidth,clip=true,trim=0.25cm 0.5cm 1cm 1.25cm]{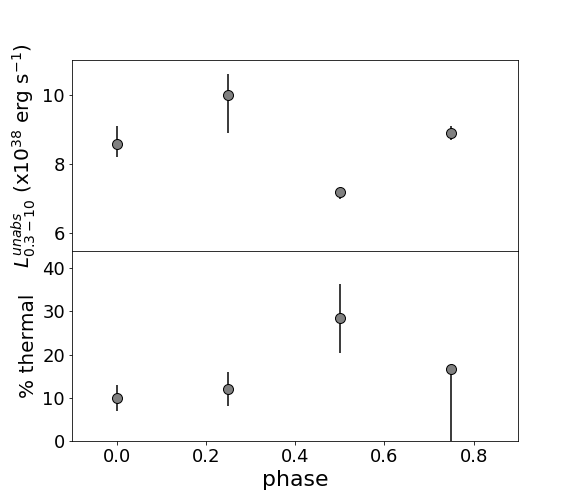}
    \caption{Average unabsorbed 0.3-10 keV \Lx\ (top) and the fraction of the luminosity originating in the thermal component (bottom) as a function of phase.}
    \label{fig:lumin_by_phase}
\end{figure}

Instead, it is the {\it unabsorbed} \Lx\ that is changing at  $\phi=0.5$ when the BH accretion disk is partially eclipsed by the WR donor, along with the fraction of the \Lx\ that originates in the thermal component (Figure~\ref{fig:lumin_by_phase}). As expected, during the eclipse, \Lx\ dips by a factor of $\sim$1.3 and the thermal fraction increases by a factor of $\sim$2.6. During non-eclipse phases, the Compton component of the X-ray spectrum accounts for $\sim$90\% of the overall value of \Lx. The thermal fraction is additionally highest during the X-ray eclipse, suggesting that it is the Comptonized component -- the BH accretion disk -- that is being eclipsed, while X-rays intrinsic to the WR star and its winds remain largely unobscured (the Comptonizing component makes up only $\sim$70\% of the overall \Lx\ at $\phi=0.50$). There is additionally evidence of an enhancement in \Lx\ at $\phi=0.25$, although no corresponding change in the proportion of thermal X-rays is observed. The average unabsorbed, 0.3-10 keV X-ray luminosity at phase $\phi=0$ is \Lx\ = $(8.6^{+0.5}_{-0.4})\times10^{38}$ \lum; \Lx\ at $\phi=0.75$ is consistent with this value.

\subsection{Discussion of Phase-Resolved X-ray Emission}
With phase-resolved spectral fitting complete, we can now convert \chandra\ and \xmm\ count rates to unabsorbed, 0.3-10 keV luminosities. We use the CIAO tool \texttt{modelflux} (along with appropriate response matrix and ancillary response files for each observation) to convert the observed count rates to flux values, using our physically-motivated spectral model and the best-fit parameters derived for each observation as a function of orbital phase. 

\begin{figure}
    \centering
    \includegraphics[width=1\linewidth]{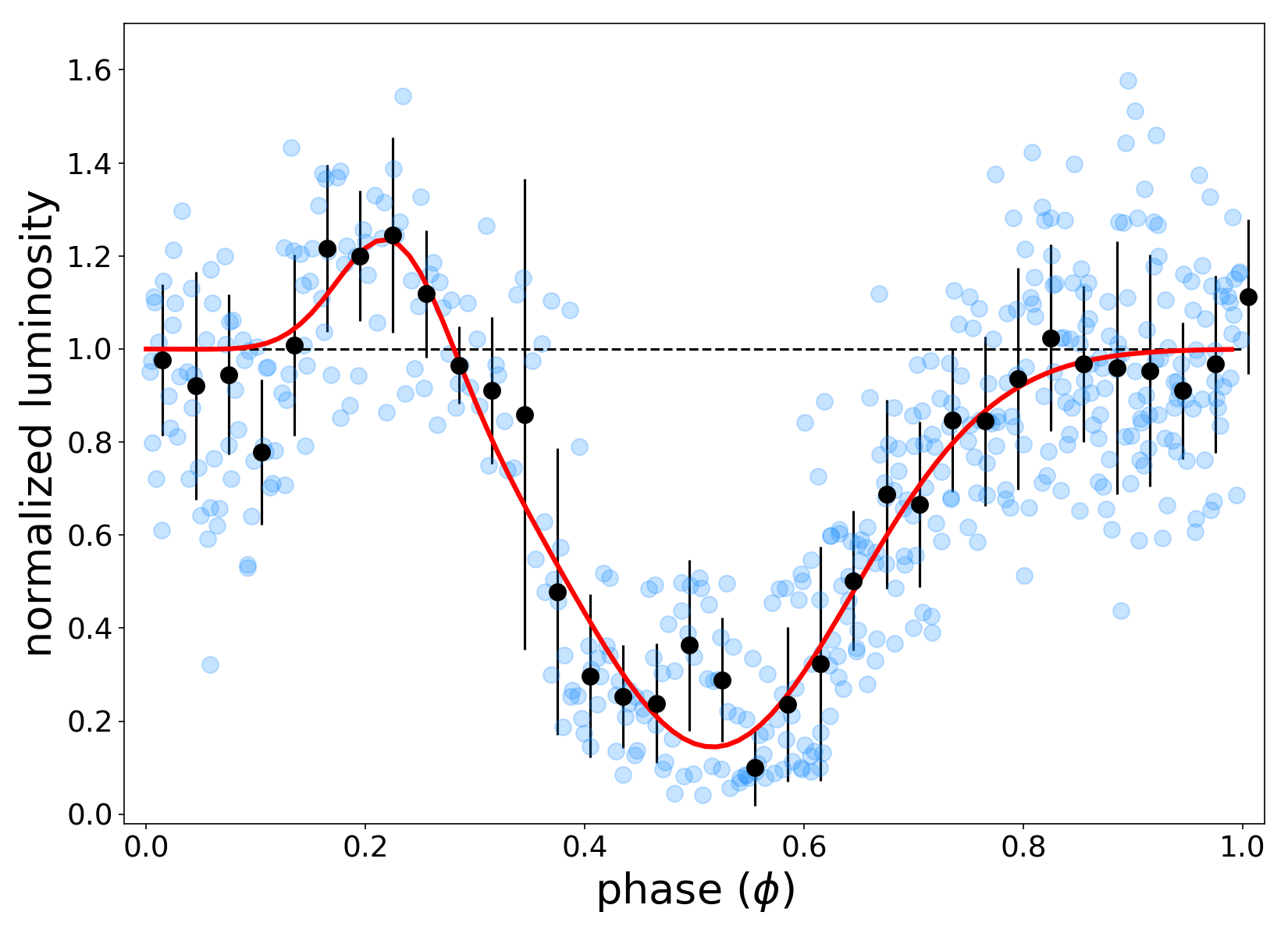}
    \caption{Reconstructed light curve with \Lx\ derived from the best-fit spectral model for each observation and orbital phase, normalized to the average \Lx\ from the $\phi=0.75$ phase ($8.6\times10^{38}$ \lum). The black points show the binned light curve. The red line shows a fit to the light curve with two Gaussian components: one representing the X-ray eclipse (centered at $\phi=0.50$), and one showing the possible accretion stream impact on the disk edge at $\phi\sim0.2$.}
    \label{fig:LC_Lnorm}
\end{figure}

The reconstructed X-ray light curve, normalized such that the average \Lx\ at $\phi=0$ is 1, is shown in Figure~\ref{fig:LC_Lnorm}. The binned light curve (black points) shows that the deepest portion of the eclipse ranges from $\phi=0.4-0.6$, and a clear \Lx\ excess is observed at $\phi\sim0.2$. We are able to fit the light curve with a double-Gaussian model: a deep, inverted Gaussian centered at $\phi\sim0.5$ corresponding to the X-ray eclipse, and a smaller Gaussian component describing the \Lx\ excess centered at $\phi=0.23$. We find a best-fit depth of the X-ray eclipse of -0.86$\pm$0.03 (i.e., a $\sim$86\% decrease in \Lx) and a best-fit amplitude of the X-ray excess of 0.30$\pm$0.07. Furthermore, there is evidence of asymmetry between the eclipse ingress and egress, with ingress occurring over $\Delta\phi\sim0.15$ and egress spanning $\Delta\phi\sim0.20$. 

Pre-eclipse humps are commonly observed (in both the optical and UV) in other accreting compact object systems, such as cataclysmic variables \citep[e.g., DW UMa][]{Hoard+10,Stanishev+04}, and have been interpreted as bright spots in the compact object's accretion disk resulting from the accretion stream from the L1 point impacting the edge of the disk. The soft X-ray photons from the shocked gas can photoionize the gas in the accretion flow and produce \ion{He}{II} lines from the vicinity of the impact site. It is therefore reasonable to assume that the pre-eclipse hump observed in Figure~\ref{fig:LC_Lnorm} is the result of an accretion stream flowing through the L1 point and impacting the outer edge of the BH accretion disk. The asymmetry in the eclipse profile further supports this interpretation; in other XRBs \citep[including the ``twin'' system  IC 10 X-1; ][]{Laycock+15a} the eclipse profile asymmetry is explained as follows \citep{Rutten+92}. Due to conservation of angular momentum, the accretion stream that impacts the outer edge of the compact object accretion disk (generating the hot spot) lags the orbital motion of the compact object. As the compact object moves into the eclipse it is partially obscured by material in the accretion stream, while during eclipse egress the compact object leads the hotspot and the accretion stream. This leads to a small deficit in the X-ray luminosity during eclipse egress, until the accretion stream and hotspot spin back into view of the observer. The temperatures and densities at these accretion stream impact sites are sufficient to produce strong \HeIIop\ emission lines \citep{Montgomery+12}, providing a plausible explanation for the phase lag observed in the \HeIIop\ RV curve \citep{Binder+15}.

\section{The Ultraviolet Spectrum} \label{sec:UVanalysis}
WR stars exhibit strong, broad emission lines \citep{Wolf+1867}, due to their dense and powerful stellar winds; the regions of the stellar wind that form these lines are geometrically extended, residing far from their parent stars, and the physical depth at which lines form within the wind is wavelength-dependent \citep{Crowther07}. Because of the physical separation between the stellar radius and the line-formation region in the wind, the stellar temperatures \Tstar\ of WR stars are difficult to measure, and non-LTE atmospheric models are needed to interpret observed spectral features and infer physical properties of the star.

Generally, models of WR atmospheres are parameterized by the luminosity of the star (\Lstar) and the stellar temperature (\Tstar). Due to the geometrical extent of the winds compared to the stellar radius, it is common to define the inner boundary \Rstar\ as the radius at which the Rosseland optical depth $\tau_{\rm R}\sim20$ \citep{Todt+15}, however the optical continuum radiation originates from the `photosphere' of the star (assumed to be where $\tau_{\rm R}\sim2/3$). In weak-lined, early-type WNs, such as the WR star in X-1 \citep[identified as a WN5 subtype;][]{Crowther+10,Crowther+07}, \Rstar\ can be on the same order as $R_{2/3}$ \citep[as in the case of HD 9974; ][]{Marchenko+04}. Optical VLT/FORS2 spectroscopy obtained by \citet{Crowther+10} has constrained \Tstar\ $\sim$ 65\,000 K ($\pm$5\,000 K), log($L$/\Lsun) $\sim$ 5.92, \massloss\ $\sim5\times10^{-6}$ \mrate, and \vinf\ $\sim$ 1\,300 \kms. These stellar parameters are consistent with a WN5 WR star with \Mstar\ = 26$^{+7}_{-5}$ \Msun, however a spectroscopic mass as low as $\sim$15 \Msun\ is possible \citep[and consistent with the masses of Galactic weak-lined WN5 stars;][]{Hamann+06}. The inferred stellar radius \Rstar\ is 7.2 \Rsun\ or 4.8 \Rsun\ for the higher- and lower-mass estimates, respectively.

Due to their high temperatures, WR stars are also bright UV sources. Correlations between specific emission lines of singly-ionized helium (\ion{He}{II}) are of particular interest, as \ion{He}{II} is a recombination line in WR winds. Specifically, \citet{Conti+90} observed an empirical correlation between the \HeIIuv\ line ($n=3\rightarrow2$, the singly-ionized helium equivalent of H$\alpha$) and \HeIIop\ ($n=4\rightarrow3$, the equivalent of Pa $\alpha$) lines in WN-type WR stars. The equivalent widths (EWs) and fluxes of the two lines are correlated for isolated WN stars; \citet{Leitherer+19} found

\begin{equation}
    \text{log}_{10} \text{EW}_{4686} = 0.97 (\text{log}_{10} \text{EW}_{1640}) + 0.58,
\end{equation}

\noindent suggesting the ratio of the two EWs is near 3.8 (and roughly independent of stellar parameters) in Galactic and LMC WN stars. The line fluxes are additionally found to be related to one another by:

\begin{equation}
    \text{log}_{10} F_{4686} = (0.95\pm0.02) \text{log}_{10}F_{1640} - (1.36\pm0.17).
\end{equation}

We obtained FUV spectra of the WR donor star in the X-1 binary system that sample the binary orbital period with \hst/COS. In this section, we present the analysis of these observations and compare them to the optical spectra (as a function of orbital phase) available in the literature \citep{Crowther+10}.

\subsection{UV Properties of the X-1 Donor Star}
To infer the spectral properties of the donor star and compare them to those derived from optical observations, we first compare our UV spectra to the Potsdam Wolf-Rayet (PoWR) \citep{Todt+15} model atmospheres. Briefly, the PoWR models iteratively solve the (non-LTE) radiative transfer equation in the comoving frame of the expanding WR atmosphere (assumping spherical symmetry) simultaneously with the equations of both radiative and statistical equilibrium \citep{Hamann+04}. The stellar radius \Rstar\ (defined as the radius at which $\tau_{\rm R}=20$), luminosity \Lstar, and temperature \Tstar\ are the main parameters of the model. Other relevant parameters are the stellar mass-loss rate \massloss\ and the terminal wind velocity \vinf, which are combined into a quantity called the ``transformed radius'' $R_{\rm t}$ \citep{Schmutz+89}:

\begin{equation}
    R_{\rm t} = R_{\star} \left[\left(\frac{v_{\infty}}{\dot{M}\sqrt{D}} \right)\left(\frac{10^{-4} ~M_{\odot}\text{ yr}^{-1}}{2\,500~\text{km s}^{-1}}\right)\right]^{2/3}.
\end{equation}

\noindent The quantity $D$ is the clumping contrast, which parameterizes the inhomogeneities in the stellar wind. There is a significant body of evidence supporting highly clumped winds in WR stars \citep{Moffat+88,Lepine+00,St-Louis+93,Kurosawa+02}, and a value of $D=10$ has been inferred for WN stars in the LMC \citep{Crowther+10b,Doran+13}. At higher (e.g. Galactic) metallicity, a value of $D=4$ is more appropriate \citep{Hamann+98}. Model spectra with equal values of $R_{\rm t}$ and \Tstar (and the same chemical composition) exhibit the same equivalent widths of emission lines. Thus, although all PoWR model grids are computed assuming log($L$/\Lsun) = 5.3, the model parameters can be scaled to match an observed luminosity (i.e., \massloss\ $\propto L^{3/4}$) or to account for a different clumping contrast (\massloss\ $\propto D^{-1/2}$).

The spectrum obtained on 2020 April 29 corresponds to a binary orbital phase $\phi=0.5$, when the WR star resides at inferior conjunction and the RV shifts due to the star's orbital motion are expected to be minimal. The brightest emission line in that of \HeIIuv, while \CIV, \ion{N}{IV} $\lambda$ 1718, and \ion{N}{IV} $\lambda$1486 (which also contains emission features from \ion{Fe}{II}) are also easily visible in the spectrum. We thus compare our observation to the PoWR models over a wavelength range of 1400--1750 \AA\ and over three metallicity regimes: metallicities typical of the Milky Way, the Large Magellanic Cloud (0.5$Z_{\odot}$), and the Small Magellanic Cloud (0.2$Z_{\odot}$). In order to directly compare our observed spectrum to the PoWR models, we bin our observed spectrum to a resolution of 0.5 \AA\ and interpolate the PoWR models to match the wavelength axis of our data. The model spectrum is then scaled to match the flux of the observed spectrum over the 1400-1750 \AA\ wavelength range. We calculate the $\chi^2$ value for each PoWR model compared to our observed spectrum, and models yielding $\chi^2<2$ (the ``good'' fits) are further inspected by eye.

\begin{figure}
\centering
\includegraphics[width=1\linewidth]{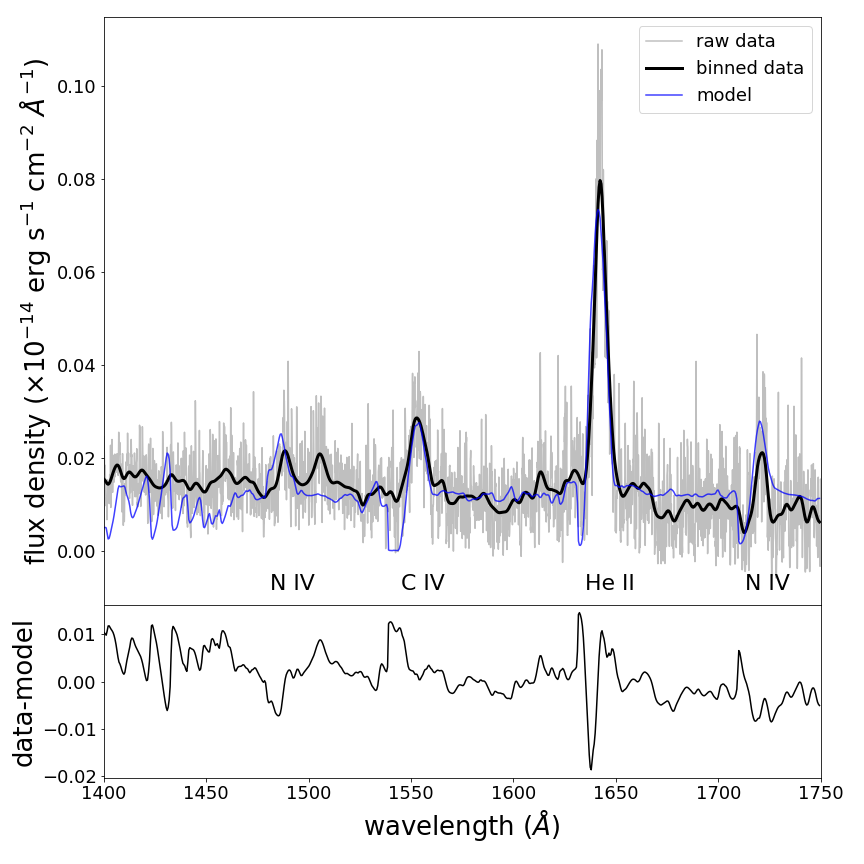} 
    \caption{{\it Top panel}: Best-fit PoWR models (blue) at LMC metallicity compared to the observed FUV spectrum (black line shows spectrum binned to 0.5 \AA\ resolution, gray shows the unbinned data). The best-fit model has \Tstar\ = 50\,100 K and $R_{\rm t}=7.85$ \Rsun. {\it Bottom panel}: data-model residuals.}
    \label{fig:PoWR_model_LMC}
\end{figure}

Early-type, hydrogen-free (WNE) models at LMC metallicity were strongly preferred over late-type (WNL) models at Galactic or SMC metallicities. This result is consistent with the measured metallicity of NGC~300 at the location of the X-1 system \citep{Urbaneja+05} and the optical spectrum, which also showed a lack of strong hydrogen features \citep{Crowther+10}. We also tested LMC-metallicity WC stellar models, but the weakness of the \CIV\ line makes the WC designation unlikely. Our observations therefore strongly support the WN designation of the donor star.

The best-fit PoWR model\footnote{Best-fit model is available at \url{http://www.astro.physik.uni-potsdam.de/~wrh/PoWR/LMC-WNE/07-12}} (which minimized the $\chi^2$ value while yielding the smallest data-model residuals) is shown in Figure~\ref{fig:PoWR_model_LMC}. It has an effective stellar temperature of 50\,100 K. We scale the best-fit model to the observed optical luminosity found by \citet{Crowther+10}, which yields $R_{\rm t}=7.85$ \Rsun\ and a mass-loss rate of \massloss\ $\sim4\times10^{-5}$ \mrate. Thus, the UV spectrum is consistent with somewhat cooler WR star than is suggested by the optical spectrum, although other models with higher values of \Tstar\ produce statistically acceptable fits to the data.

\subsection{Specific Spectral Line Measurements}
We next modeled the line profiles of specific UV lines using the Python packages \texttt{pyspeckit} \citep{Ginsburg+11} and \texttt{specutils}.\footnote{See \url{https://specutils.readthedocs.io/}} In addition to the readily observable \HeIIuv, \CIV, \ion{N}{IV} $\lambda$1718 and \ion{N}{IV} $\lambda$ 1486 features, we also detect \ion{C}{II} $\lambda$1334 in absorption. For all spectral lines, we fit a Gaussian model (or a sum of multiple Gaussian components) to the data and the observed flux in each wavelength element is then randomly shifted to a value within the flux uncertainty of the observation, and the fit is performed again. This ``bootstrapping'' is performed 5\,000 times for each spectral line, and the 16th, 50th, and 84th percentiles of the resulting parameter distributions are used to derive the best-fit value and the 90\% confidence interval of the parameters.

\subsubsection{The \HeIIuv\ Emission Line}\label{sec:HeII_RV}
\begin{figure}
\centering
\includegraphics[width=1\linewidth,clip=true,trim=0cm 1cm 0cm 0cm]{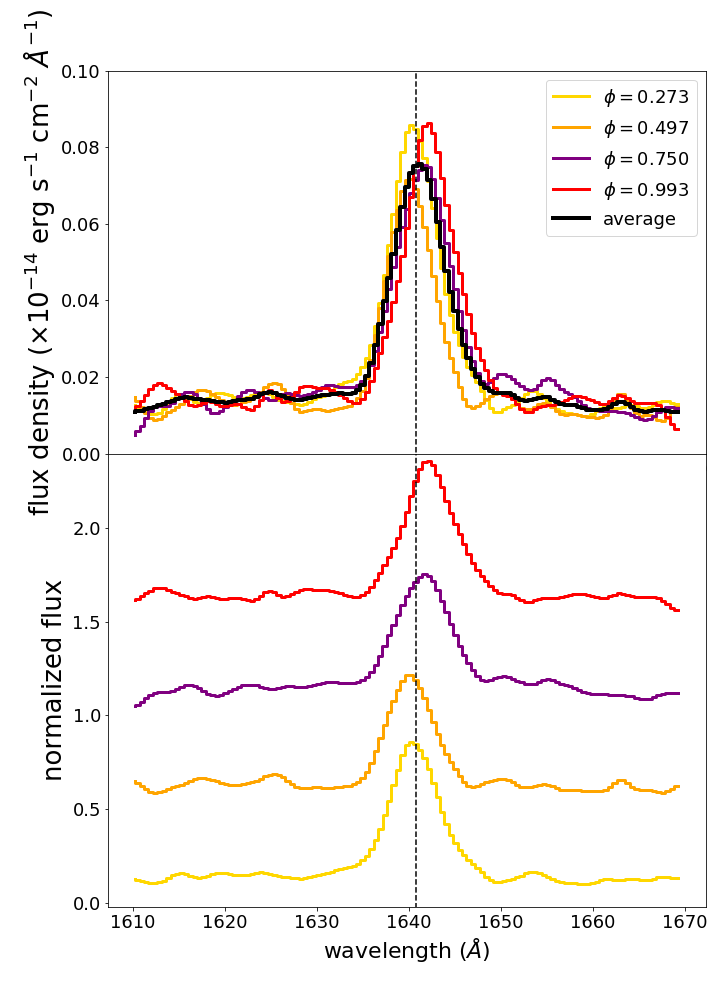} \\ 
\caption{{\it Top}: the average \ion{He}{II} $\lambda$1640 emission line (black) compared to individual observations are shown (color-coded according to orbital phase), showing a clear detection of the emission line. Spectra have been binned and smoothed for display purposes. {\it Bottom}: the \ion{He}{II} $\lambda$1640 emission line in each observation (normalized to the continuum level), with the best-fit Gaussian superimposed (the fluxes at each phase $\phi$ have an arbitrary offset applied for clarity). In both panels, the dashed line shows the rest-frame wavelength of \ion{He}{II} $\lambda$1640, corrected for the recessional velocity of NGC~300.}
\label{fig:HeII1640}
\end{figure}

\HeIIuv\ is by far the strongest emission line present in the WR spectrum (see Figure~\ref{fig:HeII1640}). The best-fit parameters for the emission line are summarized in Table~\ref{tab:HeII_spec_line}. In WR star winds, \ion{He}{II}\ is excited collisionally, and is a hallmark of the WN subtype. However, \ion{He}{II}\ emission lines can also arise in the hot spots of compact object accretion disks \citep{Montgomery+12,Steeghs+02,Cowley+75} or colliding wind shocks \citep[as has been studied extensively for $\eta$ Car;][]{Martin+06,Mehner+11,Teodoro+12,Madura+13,Teodoro+16}, and thus the interpretation of this feature in a known HMXB requires some caution. In the optical, the \HeIIop\ EW is $\sim$56 \AA\footnote{Typically positive values of EW indicate an absorption line, however throughout this work we refer to EWs in terms of their absolute values.} and FWHM is $\sim$17\AA, somewhat lower than observed in LMC and Milky Way WN stars \citep{Crowther+10}. Given the observed correlation between the optical and UV EWs found by \citet{Leitherer+19}, we therefore may expect the \HeIIuv\ EW to be $\approx$15 \AA.

\begin{deluxetable*}{ccccc}
\tablenum{4}
\tablecaption{\ion{He}{II} $\lambda$1640 Line Measurements\label{tab:HeII_spec_line}}
\tablewidth{0pt}
\tablehead{
& \multicolumn{4}{c}{Orbital Phase} \\ \cline{2-5}
\colhead{Parameter} & \colhead{0.273} & \colhead{0.497} & \colhead{0.750} & \colhead{0.993}  
}
\decimalcolnumbers
\startdata
$\lambda_{\rm c}$ (\AA)     & 1640.89$\pm$0.16  & 1640.42$\pm$0.15  & 1641.73$^{+0.20}_{-0.19}$ & 1642.73$^{+0.17}_{-0.18}$ \\
RV$_{\rm cor}$ (\kms)   & -60$\pm$30 & -150$\pm$30 & 90$\pm$40 & 270$\pm$30 \\
EW (\AA)                &  28.4$\pm$1.2 & 24.6$^{+0.8}_{-0.9}$ & 27.2$^{+1.5}_{-1.4}$ & 29.8$\pm$0.9  \\
FWHM (\AA)            & 6.14$\pm$0.25 & 4.65$^{+0.15}_{-0.16}$ & 5.95$^{+0.33}_{-0.31}$  & 5.93$^{+0.17}_{-0.18}$ \\
FWHM (\kms)           & 2245$\pm$90 & 1700$^{+55}_{-59}$ & 2175$^{+121}_{-113}$ & 2170$^{+62}_{-66}$  \\
$F_{\rm line}$ ($\times10^{-15}$ erg s$^{-1}$ cm$^{-2}$) & 4.67$\pm$0.04 & 3.43$\pm$0.03 & 4.14$\pm$0.04 & 4.93$\pm$0.04 \\
log$L_{\rm line}$ ([erg s$^{-1}$]) & 36.35$^{+0.03}_{-0.04}$ & 36.22$\pm$0.04 & 36.30$^{+0.04}_{-0.05}$ & 36.37$\pm$0.04 \\
\enddata
\end{deluxetable*}

Instead, Table~\ref{tab:HeII_spec_line} reveals an emission line that exhibits strong variability with orbital phase. At $\phi=0.497$, when the BH accretion disk is largely eclipsed and we expect to view the WR winds with minimal X-ray irradiation, the \HeIIuv\ EW is $\sim$25 \AA, and the FWHM and line fluxes are both at minimum values. This phase also corresponds to the strongest observed blueshift of the \HeIIuv\ line (note that the radial velocities, RV$_{\rm cor}$, listed in Table~\ref{tab:HeII_spec_line} have been corrected for the recessional velocity of NGC~300). Approximately half an orbital period later, at $\phi=0.993$, we observe a strong redshift, a maximum line flux, and slightly larger values of the EW ($\sim$20\% larger than at $\phi=0.497$) and FWHM ($\sim$30\% higher than at $\phi=0.497$). The EW$_{1640}$ is about twice the value predicted for single WN stars \citep{Leitherer+19}.

\begin{figure}
     \includegraphics[width=1\linewidth,clip=true,trim=0.5cm 0.5cm 0.5cm 1.75cm]{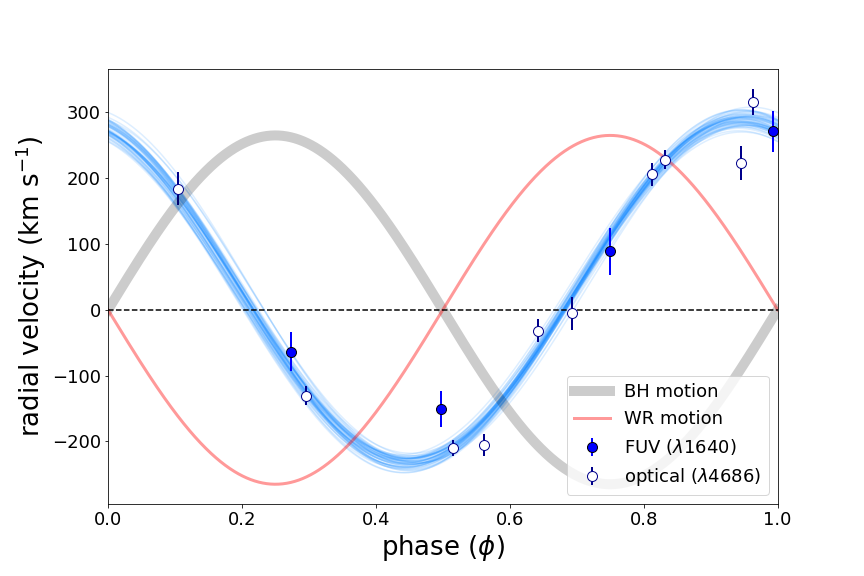} 
\caption{The radial velocity curves for \HeIIuv\ (solid blue circles) and \HeIIop\ \citep[open circles, taken from][]{Crowther+10}. The expected BH orbital motion (scaled to approximately match the amplitude of the \ion{He}{II} curve) is shown for reference in gray, and the expected WR orbital motion is shown in red. The light blue lines show 50 random draws from our fits to the data.}
\label{fig:HeII_phased}
\end{figure}

We use our updated ephemeris from Section~\ref{sec:period} to re-compute the phases at which the optical \HeIIop\ RVs from \citet{Crowther+10} were obtained. We combine these with our \HeIIuv\ measurements and fit a sinusoid to the data (we perform 500 iterations of the fit, randomly adjusting the RV measurements within the observed uncertainties, and again report the median and 90\% confidence interval of the resulting fit parameter distributions). The results are shown in Figure~\ref{fig:HeII_phased}. We measure the RV semi-amplitude $K$ and the phase-shift $\Delta \phi$ from the expected BH orbital motion. The results are summarized in Table~\ref{tab:HeII_RV_fit}. In addition to the combined UV + optical data set, we also fit the UV and optical measurements independently \citep[our optical-only value of $K$ is in excellent agreement with][]{Crowther+10}.

There is a systematic phase delay of $\Delta \phi=0.30$ observed in the \ion{He}{II}\ RV data, compared to the expected orbital motion of the BH. There is additionally marginal evidence (at approximately $2\sigma$ significance) for different values of $K$ between the optical and UV data sets, with the UV observations showing a lower RV amplitude, suggesting that the two lines may be originating from slightly different locations in the X-1 system. When taken together, the observed phase delay and variable line profile indicate that bright \ion{He}{II} emission (or at least a significant component of the line emission) is not originating uniformly in the WR wind, but is instead arising from a more complex interaction between the two binary components. We discuss possible interpretations of the observed behavior in more detail in Section~\ref{sec:discussion}.

\begin{deluxetable}{ccc}
\tablenum{5}
\tablecaption{\ion{He}{II}\ Radial Velocity Curve Fit Parameters\label{tab:HeII_RV_fit}}
\tablehead{
\colhead{Data Set} & \colhead{$K$ (\kms)} & \colhead{$\Delta \phi$} }
\decimalcolnumbers
\startdata
UV + optical   & 260$\pm$10 & 0.30$\pm$0.01 \\
UV only        & 220$\pm$20 & 0.30$\pm$0.02 \\
optical only    & 270$\pm$10 & 0.30$\pm$0.01 \\
\enddata
\end{deluxetable}

\subsubsection{Fainter Emission Lines}\label{sec:CIV_RV}
\begin{figure}
\centering
\includegraphics[width=1\linewidth]{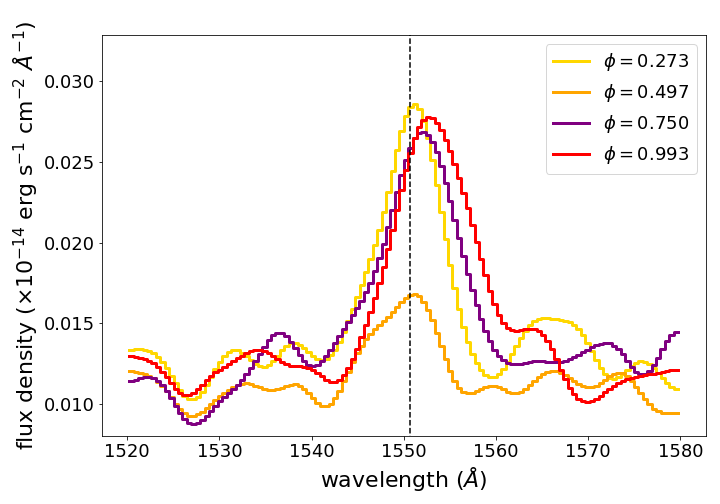}
\caption{The \CIV\ spectral feature line in each observation, smoothed for display purposes. The dashed line shows the rest-frame wavelength corrected for the recessional velocity of NGC~300.}
    \label{fig:CIV1550}
\end{figure}

The second strongest line detected in the \hst/COS spectra is the UV resonance line \CIV\ (shown in Figure~\ref{fig:CIV1550}, roughly a factor of four fainter than \HeIIuv\ in flux), which is often used to estimate \vinf\ of WR stars. In addition to the emission peak at 1550 \AA, the P Cygni absorption component at 1548 \AA\ is typically seen in WR stars (and is present in the PoWR model). The depth and width of the \CIV\ doublet are sensitive to the inclusion of X-ray radiation in the WR model atmosphere \citep{Gimenez-Garcia+16}. Although there are hints of a P Cygni line profile during phases 0.750 and 0.993, the signal-to-noise ratio is not sufficient for us to directly model multiple components; a single component Gaussian generally matches the largest flux value in amplitude, and the FWHM is calculated independently of the Gaussian model. Like \HeIIuv, the \CIV\ (see Table~\ref{tab:CIV_spec_line}) is clearly complex and variable, with EW, FWHM, and line fluxes that vary as a function of orbital phase. 

\begin{deluxetable*}{ccccc}
\tablenum{6}
\tablecaption{\ion{C}{IV} $\lambda$1550 Line Measurements\label{tab:CIV_spec_line}}
\tablewidth{0pt}
\tablehead{
& \multicolumn{4}{c}{Orbital Phase} \\ \cline{2-5}
\colhead{Parameter} & \colhead{0.273} & \colhead{0.497} & \colhead{0.750} & \colhead{0.993}  
}
\decimalcolnumbers
\startdata
$\lambda_{\rm c}$ (\AA)     & 1551.00$\pm$0.08  & 1551.44$\pm$0.04  & 1553.56$^{+0.08}_{-0.32}$ & 1551.96$\pm$0.08 \\
RV$_{\rm cor}$ (\kms)   & -100$\pm$30 & -10$\pm$20 & 400$^{+30}_{-60}$ & 90$\pm$30 \\
EW (\AA)                &  10.0$\pm$0.1 & 5.3$^{+0.6}_{-0.7}$ & 9.8$^{+0.9}_{-0.8}$ & 13.9$\pm$0.1  \\
FWHM (\AA)            & 2.8$\pm$0.5 & 1.9$\pm$0.1 & 6.3$\pm$0.1  & 6.1$\pm$0.1 \\
$F_{\rm line}$ ($\times10^{-15}$ erg s$^{-1}$ cm$^{-2}$) & 1.15$\pm$0.01 & 0.46$\pm$0.01 & 1.25$\pm$0.01 & 1.44$\pm$0.01 \\
\enddata
\end{deluxetable*}

\begin{figure}
\includegraphics[width=1\linewidth]{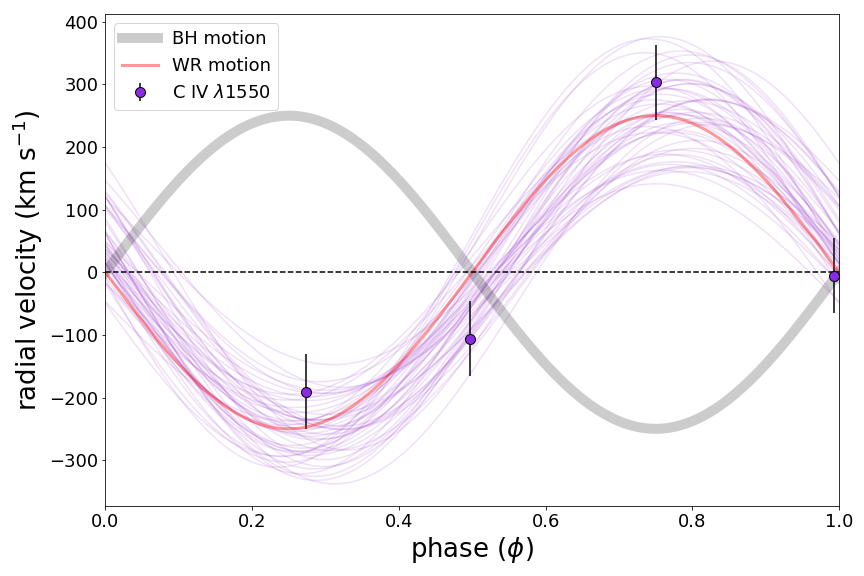}
\caption{Same as Figure~\ref{fig:HeII_phased}, but for \ion{C}{IV} $\lambda$1550.}
\label{fig:CIV_phased}
\end{figure}

RV variations with orbital phase are also apparent in the line profile. The RV curve of \CIV\ is shown in Figure~\ref{fig:CIV_phased}. The general shape of the \CIV\ RV curve matches the expected orbital motion of the WR star. We assume the \CIV\ RVs follow a sinusoidal curve (although with only four phase samples available, we cannot rigorously demonstrate that this is the case) and follow the same fitting procedure as for \HeIIuv. A systematic redshift of roughly 95 \kms\ is observed, with a semi-amplitude $K$ = 250$\pm$40 \kms\ and a phase offset consistent with zero ($\Delta \phi$ = 0.03$\pm$0.03). 

Other lines present in the WR spectrum are weaker still than \CIV. Their properties are summarized in Table~\ref{tab:faint_lines}. Given their low fluxes (an order of magnitude below \CIV) and complex line profiles (for example, \ion{N}{IV} $\lambda$1486 is a blend of several \ion{N}{IV} and \ion{Fe}{V} lines), we do not attempt to fully model these lines. 

\begin{deluxetable}{cccccc}
\tablenum{7}
\tablecaption{Properties of Other Faint Spectral Lines\label{tab:faint_lines}}
\tablewidth{0pt}
\tablehead{
 & \colhead{Orbital} &  \colhead{$\lambda_C$} & \colhead{FWHM} & \colhead{EW} & \colhead{$F_{\rm line}^a$}  \\
\colhead{line ID} & \colhead{Phase}   & \colhead{(\AA)}   & \colhead{(\AA)}   & \colhead{(\AA)}   & \colhead{(erg s$^{-1}$ cm$^{-2}$)}}
\decimalcolnumbers
\startdata
                & 0.273 & 1335.08 & 1.4 & -1.2 & \nodata \\
\ion{C}{II}    & 0.497 & 1334.84 & 2.0 & -1.0 & \nodata \\
$\lambda$1334   & 0.750 & 1334.84 & 2.2 & -0.5 & \nodata \\
                & 0.993 & 1335.08 & 1.6 & -1.0 & \nodata \\
\hline
                & 0.273 & 1486.86 & 5.8 & 6.0 & 7.2$\times10^{-16}$ \\
\ion{N}{IV}     & 0.497 & 1484.48 & 6.2 & 5.2 & 5.9$\times10^{-16}$ \\
$\lambda$1468   & 0.750 & 1486.13 & 4.8 & 2.9 & 3.9$\times10^{-16}$ \\
                & 0.993 & 1488.30 & 5.1 & 1.4 & 1.3$\times10^{-16}$ \\
\hline
                & 0.273 & 1718.15 & 5.7 & 3.7 & 4.8$\times10^{-16}$ \\
\ion{N}{IV}     & 0.497 & 1721.03 & 7.7 & 11.2 & 5.4$\times10^{-16}$ \\
$\lambda$1718   & 0.750 & 1720.06 & 5.6 & 5.4 & 6.0$\times10^{-16}$ \\
                & 0.993 & 1721.50 & 5.6 & 5.8 & 5.5$\times10^{-16}$ \\
\enddata
\tablecomments{$^a$Fluxes are not reported for absorption lines, and we adopt a convention where the absorption line EWs are negative.}
\end{deluxetable}

\section{Discussion\label{sec:discussion}}
The major results of our joint \chandra-\hst\ program (supplemented with archival X-ray observations) include the pre-eclipse hump in the X-ray light curve, the phase delay in the optical and UV \ion{He}{II} line RVs, and the RV variations observed in the \CIV\ emission line. In this section, we discuss these results in more detail (including necessary assumptions and caveats), and present a model for the accretion flow in the X-1 binary system.

\subsection{Binary Parameters and Black Hole Mass}\label{section:BHmass}
We can use the refined orbital period and \CIV\ RV semi-amplitude (assuming the RV variations follow a sinusoidal curve) to estimate the binary parameters and, ultimately, constrain the mass of the X-1 BH. \citet{Crowther+10} constrain the inclination of the X-1 system to $i=60-75^{\circ}$, and our light curve (Figure~\ref{fig:LC_Lnorm}) is consistent with these values. 

In Section~\ref{sec:CIV_RV}, we discovered the RV of the \CIV\ emission feature closely tracks the expected orbital motion of the WR star, with a semi-amplitude $K$ = 250$\pm$40 \kms. This measurement allows us to derive an updated mass estimate of the BH:

\begin{equation}
    M_{\rm BH} = \frac{P K^3}{2 \pi G} \frac{(1+q)^2}{\text{sin}^3 i}
\end{equation}

\noindent where $q$ is the mass ratio between the donor star (\Mstar) and the compact object ($q$ = \Mstar/\MBH). We assume the preferred \Mstar\ = 26 \Msun\ from \citet{Crowther+10} and use our measurements of the orbital period ($P=32.7921$ hr), and $K$ from the \CIV\ line. For inclination angles ranging from 60$^{\circ}$-75$^{\circ}$, we derive \MBH = 17$\pm$4 \Msun. This is somewhat lower than the original estimate of $\sim$20 \Msun. If we assume \Mstar\ = 15 \Msun, which is also consistent with the \citet{Crowther+10} results (and is a typical value of WN stars in the Milky Way), the BH mass is reduced to 13 \Msun.

Using our new estimate of \MBH\ = 17 \Msun, we can now utilize Kepler's third law to solve for the orbital separation $a$:

\begin{equation}
    a = \left[ \frac{G\left(M_{\rm BH} + M_{\star}\right)}{4 \pi^2} P^2\right]^{1/3}
\end{equation}

\noindent and estimate the size of the WR Roche lobe \citep{Eggleton+83}:

\begin{equation}
    r_{\rm L} = \left[ \frac{0.49 q^{2/3}}{0.6q^{2/3} + \text{ln}\left(1+q^{1/3}\right)} \right] a.
\end{equation}

\noindent We find an orbital separation $a=18.2^{+0.5}_{-0.6}$ \Rsun\ and $r_{\rm L}=7.6$ \Rsun\ (implying an orbital speed of $\sim$670 \kms). Assuming a \Rstar = 7.2 \Rsun, we estimate \Rstar/$a$ = 0.4 and \Rstar/$r_{\rm L}$ = 0.95. If we instead assume a lower-mass system (with \Mstar\ = 15 \Msun, \Rstar\ = 4.8 \Rsun, and \MBH\ = 12 \Msun), we find $a=15.8$ \Rsun\ and $r_{\rm L}=6.2$ \Rsun, implying \Rstar/$a$ = 0.3 and \Rstar/$r_{\rm L}$ = 0.77. Thus, the WR star does not completely fill its Roche lobe, making it unlikely that the standard model of Roche lobe overflow (RLOF) is occurring in the X-1 system.

\subsection{The Mass Accretion Geometry}
Without standard RLOF, mass transfer in the NGC~300 X-1 system must occur through gravitational interactions between the BH and the WR winds via a process similar to Bondi-Holye-Lyttleton wind accretion \citep[BHL; ][]{Bondi+44,Edgar+04}. This standard model of direct wind accretion typically yields X-ray luminosities much lower than those observed in X-1 \citep[by $\sim$2-3 orders of magnitude; see also][and references therein]{MartinezNunez+17}. More recent simulations of wind-capture disks \citep{Huarte-Espinosa+13} and wind-RLOF \citep{ElMellah+19a,ElMellah+19b}, however, provide a mechanism by which a compact object may form an accretion disk with mass accretion rate sufficient to maintain the high \Lx\ without the need for standard RLOF from the donor star (the relevant Roche lobe in the wind-RLOF scenario is that of the {\it compact object}, not of the donor star). Indeed, \citet{Tutukov+16} invoked such an explanation to reconcile the high \massloss\ values of WR donors and the observed \Lx\ of both NGC~300 X-1 and IC~10 X-1.

When the winds of the donor star are slow relative to the orbital motion of the compact object, the gravitational influence of the compact object can strongly beam the wind towards the accretor. Such a scenario has been explored for symbiotic binaries \citep{Podsiadlowski+07}, and its application to ultraluminous sources (ULXs) and supergiant XRBs (SgXBs) is an active area of study \citep{Huarte-Espinosa+13,ElMellah+19a,ElMellah+19b}. The terminal speeds of WR winds can be slowed considerably when irradiated by a strong X-ray source \citep{Hatchett+77,Fransson+80,Gimenez-Garcia+16}. This ionization of the wind is expected to occur even for relatively large separations between the donor star and the X-ray source \citep{Blondin+91,Manousakis+15}. This beamed, slowly-moving wind material does not directly accrete onto the compact object, but forms a downstream wake and flows into an accretion stream as shown in Figure 1 of \citet{Huarte-Espinosa+13} or Figure 4 of \citet{ElMellah+19a}. Due to the orbital motion of the binary components, the focused wind material is accelerated towards the earlier position of the accretor, and so the final impact of material flowing through the accretion stream and the accretion disk does not occur on the portion of the disk facing the donor star.

\begin{figure*}
    \centering
    \includegraphics[width=1\linewidth,clip=true,trim=3cm 3cm 3cm 3cm]{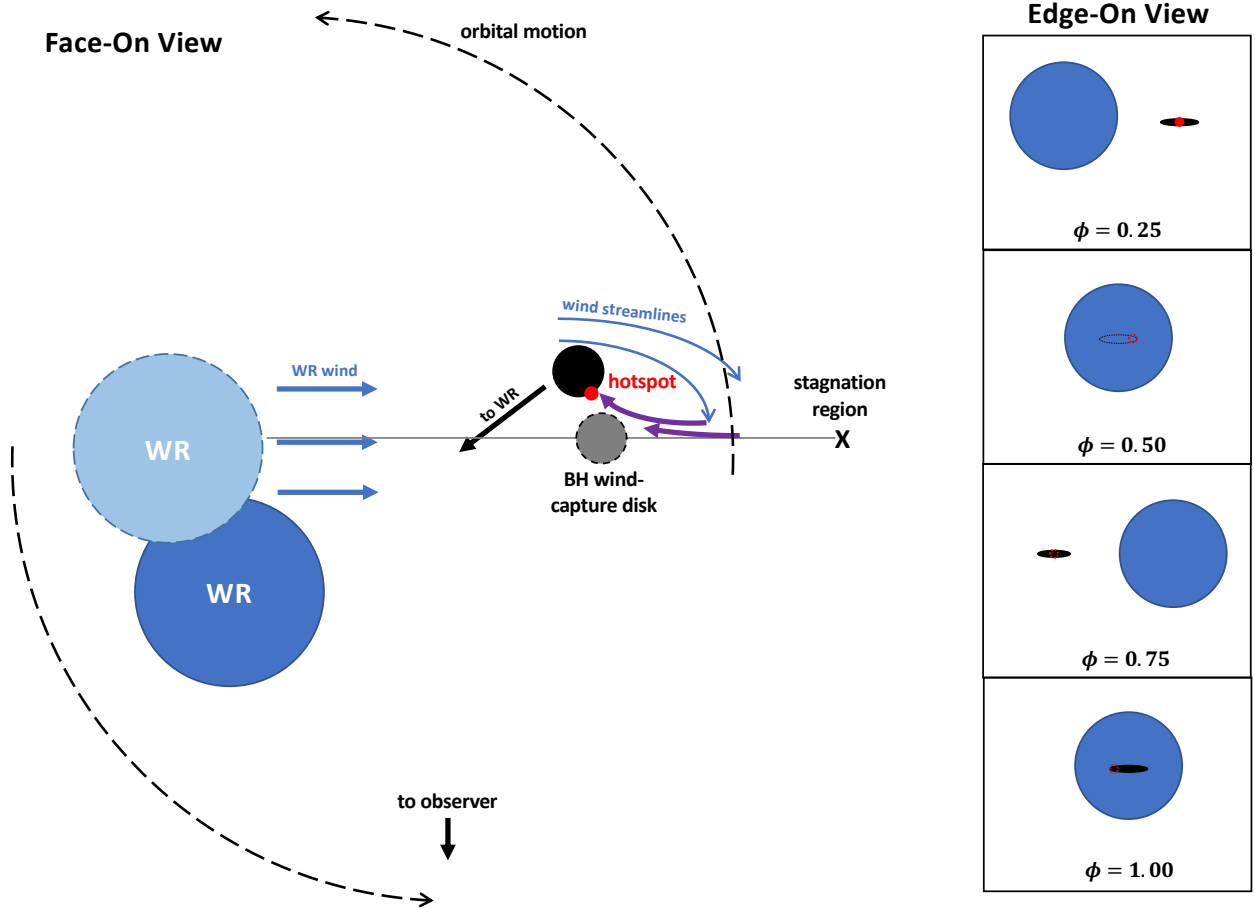}
     \caption{Toy model of the X-1 geometry, not to scale. The left side of the figure presents a face-on view of the system. The WR star is shown in blue, and the blue arrows indicate the flow of the WR winds. The black circle indicates the BH accretion disk. The lighter, dashed-outline circles show the prior locations of the binary components due to their orbital motion. The purple arrows indicate the formation of the accretion stream, which then impacts the BH disk and generates a hot spot (red). The right side of the figure shows the edge-on view of the system. Filled shapes indicate a component is visible to the observer; dotted outlines indicate the structure is largely or entirely obscured from the observer's line of sight. Observations of X-1 imply a inclination of $i=60-75^{\circ}$.}
    \label{fig:toy_model}
\end{figure*}

Not only does this model reproduce the X-ray luminosities observed in X-1, it also provides a natural mechanism by which a hot spot on the accretion disk can form at $\phi\sim0.2$. In Figure~\ref{fig:toy_model}, we present a ``cartoon model'' of the X-1 system inspired by the hydrodynamical simulations of \citet{ElMellah+19a,ElMellah+19b} and \citet{Huarte-Espinosa+13} that illustrates how the wind-capture model may give rise to an accretion disk with a hot spot consistent with our observations. The WR star is shown in blue, with blue arrows indicating the flow of the WR winds (not to scale). The BH accretion disk is shown in black. The purple arrows show the formation of the accretion stream, which impacts the disk and generates a hot spot (red). When viewed edge-on, the hot spot is visible to an observer at phases $\phi\sim0.25$. Neither the BH accretion disk nor the hot spot is visible during the eclipse ($\phi=0.5$), and when the accretion disk is visible at phases $\phi\sim0.75-1.0$ the hot spot is on the side of the accretion disk opposite to the observer. 

There is an apparent dichotomy observed among stellar-mass BHs in XRBs: wind-fed BHs (notably Cyg~X-1, LMC~X-1, and M33~X-7) exhibit systematically higher masses and higher BH spins compared to BHs fed via RLOF \citep{McClintock+14,Ozel+10,Steiner+14,Fragos+15}. This dichotomy has been attributed to the age of the system: BHs in wind-fed systems contain massive companions and therefore must be young, while the majority of BHs undergoing RLOF are observed in low-mass XRBs and are thus likely to be much older. The emerging picture of NGC~300 X-1 as a high-mass stellar BH powered via a wind-captured disk is thus consistent with observations of other comparable systems.

\subsection{The Origin of the \ion{He}{II} and \ion{C}{IV} Emission Lines}
The model described above, with a wind-captured accretion disk being impacted by an accretion stream originating in an X-ray irradiated wind, can additionally be used to interpret the prominent emission lines discussed in Section~\ref{sec:UVanalysis}: \HeIIuv\ (and \HeIIop\ from the literature) and \CIV.

The changes in the \HeIIuv\ line luminosity are perhaps the least surprising: if a significant portion of the accretion disk is being eclipsed by the WR star at $\phi=0.5$, we expect the UV emission to be similarly diminished. Indeed, this is what we observe, with the line luminosity increasing to a maximum value at phases $\sim$0-0.25, presumably as the accretion disk and impact site hot spot are rotating back into full view of the observer. The inclination of the system is not perfectly edge-on, as depicted in Figure~\ref{fig:toy_model}, so we would not expect the accretion disk's contribution to the UV emission to vanish (just as the X-ray emission does not vanish at $\phi$=0.5). Furthermore, both the UV and optical \ion{He}{II} emission lines could be formed in shocked regions within the accretion stream, which can have significant physical extent orthogonal to the plane of the accretion disk \citep{ElMellah+19a}. Because the accretion stream itself can have a large physical extent from the BH accretion disk (both in the plane of the binary as well as orthogonal to the orbital plane), the phase lag observed in the \ion{He}{II} may indicate that these lines form in a hot, dense portion of the accretion stream that lags the black hole.

A similar decrease in both line luminosity and EW at $\phi=0.5$ are also seen in the \CIV\ line, despite the RV variations that coincide very strongly with the expected motion of the WR star's orbital motion. The fact that the \CIV\ line flux varies with orbital phase (specifically, the decline in the line flux near $\phi=0.5$) suggests that there is an asymmetric component to the line flux. This is likely due to X-ray irradiation of the wind by the BH accretion disk: X-ray irradiation significantly enhances the \ion{C}{IV} population in the wind \citep[as in Vela X-1;][]{Gimenez-Garcia+16} and decreases the terminal velocity of the irradiated wind \citep{Hatchett+77,Fransson+80}. As a resonance line, the blue \ion{C}{IV} $\lambda$1548 line experiences a larger optical depth to resonant scattering than the red $\lambda$1550 line (with relative oscillator strengths of $f_{1548}/f_{1550}\sim2$). As the optical depth increases, the 1548 \AA\ photons are scattered out of the line of sight. Given that no strong feature was detected blueward of the 1550 \AA\ emission line, we can infer that the \CIV\ emission line originates in an optically thick portion of the WR wind, where it is excited primarily via collisions. Therefore, during the eclipse we are likely seeing only the \CIV\ emission originating in the portion of the WR wind that is ``shadowed'' from the X-rays, while at other orbital phases we can observe the contribution from \CIV\ emission originating in the photoionized portion of the WR wind. As recombination lines, the optical and UV \ion{He}{II} lines form predominantly in the lower density, lower optical depth regions of the X-1 system -- such as the accretion stream and BH accretion disk hotspot.

In using the \CIV\ emission line to trace the RV of the donor star, we have assumed that the emission-weighted mean line velocity is the same as the center of mass velocity of the WR star. This assumption may not be valid if significant asymmetries exist in the \CIV\ line as a function of phase, or if the terminal speed of the WR wind varies in different directions due to the X-ray irradiation from the BH accretion disk. Similar issues are discussed in detail \citet{Koljonen+17}, where phase-resolved optical and IR spectroscopy of Cyg~X-3 was analyzed. To fully evaluate the extent to which these assumptions affect our BH mass estimate would require a significantly larger investment of \hst\ observing time or more detailed theoretical modeling of the WR wind structure, which is beyond the scope of this work. Comparing NGC~300 X-1 to other similar systems (e.g., such as Cyg X-3 or IC~10 X-1) can also aid resolving these issues.

\section{Summary}\label{sec:summary}
We have presented an analysis of new \chandra/ACIS-I and \hst/COS observations of the Wolf-Rayet + black hole binary NGC~300 X-1. The combination of our new \chandra\ exposure with nearly two decades of archival X-ray observations with \chandra\ and \xmm\ have yielded a highly precise value of the binary orbital period: 32.7921$\pm$0.0003 hr. Phase-resolved X-ray spectroscopy indicates that the underlying X-ray continuum of X-1 remains constant throughout the binary's orbital period, with the unabsorbed X-ray luminosity decreasing during a partial eclipse by the WR star. We find that $\sim$70\% of \Lx\ originated from a Comptonizing component, likely associated with the black hole and its accretion disk, with the remainder originating in a soft, thermal component that may be due to shocks both within the WR winds or the collision between the WR winds and BH accretion disk winds. The X-ray light curve shows evidence of a pre-eclipse hump, which we interpret as an accretion stream impacting the edge of the BH accretion disk (analogous to the accretion streams seen in the UV light curves of some CVs).

The UV spectrum is generally consistent with models of WN stars (at LMC-metallicity) from the PoWR simulations with \Tstar\ $\sim$50 kK and a mass-loss rate of \massloss\ $\sim4\times10^{-5}$ \mrate. The brightest emission line in the UV spectrum belongs to \HeIIuv, which shows distinct radial velocity variations over the binary orbital period. However, these RV shifts exhibit a systematic offset from the predicted motion of the binary components. Just as with the optical \HeIIop\ emission lines, \HeIIuv\ lags the expected BH motion by $\Delta \phi=0.30$. The equivalent width is also larger than anticipated for a WN stars, indicating that these \ion{He}{II} emission lines are primarily originating from the non-stellar component of the system. \CIV\ is found to closely track the expected WR orbital motion. Assuming the \CIV\ RV variations follow a sinusoidal curve, we measure an RV semi-amplitude of 250$\pm$40 \kms, implying a BH mass of \MBH\ = $17\pm4$ \Msun\ (assuming a WR mass of 26 \Msun) and an orbital separation of $\sim$18.2 \Rsun\ between the WR and BH.

The derived binary parameters indicate that the WR star does not completely fill its Roche lobe. We therefore favor a wind-captured accretion disk scenario, where material from the WR winds is gravitationally focused into a thin disk around the BH. Under this model, wind material is funneled into an accretion stream that may be significantly misaligned with the orbital motion (and orbital plane) of the system, explaining the observed phase lags in the \ion{He}{II} emission lines and the existence of a hot spot where the accretion stream impacts the edge of the disk.

\acknowledgments
We thank the anonymous reviewer for comments and suggestions that improved this manuscript. Support for this work was provided by the National Aeronautics and Space Administration through \chandra\ Award Number GO0-21031X issued by the \chandra\ X-ray Observatory Center, which is operated by the Smithsonian Astrophysical Observatory for and on behalf of the National Aeronautics Space Administration under contract NAS8-03060. D.M.C. and S.L. acknowledge support from the NASA Astrophysics Data Analysis Program grant 80NSSC18K0430. This research has made use of data obtained from the \chandra\ Data Archive and software provided by the \chandra\ X-ray Center (CXC) in the application package CIAO. This research is based on observations made with the NASA/ESA \hst\ obtained from the Space Telescope Science Institute, which is operated by the Association of Universities for Research in Astronomy, Inc., under NASA contract NAS 5–26555. These observations are associated with program 15999. This work is based on observations obtained with \xmm, an ESA science mission with instruments and contributions directly funded by ESA Member States and NASA. This research made use of Astropy, a community-developed core Python package for Astronomy \citep{astropy:2013, astropy:2018}.

\vspace{5mm}
\facilities{HST(COS), CXO, XMM-Newton}

\software{astropy \citep{astropy:2013,astropy:2018}, pyspeckit \citep{Ginsburg+11}, CIAO \citep{Fruscione+06}}

\bibliography{sample63}

\begin{thebibliography}{}
\expandafter\ifx\csname natexlab\endcsname\relax\def\natexlab#1{#1}\fi
\providecommand{\url}[1]{\href{#1}{#1}}
\providecommand{\dodoi}[1]{doi:~\href{http://doi.org/#1}{\nolinkurl{#1}}}
\providecommand{\doeprint}[1]{\href{http://ascl.net/#1}{\nolinkurl{http://ascl.net/#1}}}
\providecommand{\doarXiv}[1]{\href{https://arxiv.org/abs/#1}{\nolinkurl{https://arxiv.org/abs/#1}}}

\bibitem[{{Aftab} {et~al.}(2019){Aftab}, {Paul}, \& {Kretschmar}}]{Aftab+19}
{Aftab}, N., {Paul}, B., \& {Kretschmar}, P. 2019, \apjs, 243, 29,
  \dodoi{10.3847/1538-4365/ab2a77}

\bibitem[{{Arnaud}(1996)}]{Arnaud96}
{Arnaud}, K.~A. 1996, in Astronomical Society of the Pacific Conference Series,
  Vol. 101, Astronomical Data Analysis Software and Systems V, ed. G.~H.
  {Jacoby} \& J.~{Barnes}, 17

\bibitem[{{Astropy Collaboration} {et~al.}(2013){Astropy Collaboration},
  {Robitaille}, {Tollerud}, {Greenfield}, {Droettboom}, {Bray}, {Aldcroft},
  {Davis}, {Ginsburg}, {Price-Whelan}, {Kerzendorf}, {Conley}, {Crighton},
  {Barbary}, {Muna}, {Ferguson}, {Grollier}, {Parikh}, {Nair}, {Unther},
  {Deil}, {Woillez}, {Conseil}, {Kramer}, {Turner}, {Singer}, {Fox}, {Weaver},
  {Zabalza}, {Edwards}, {Azalee Bostroem}, {Burke}, {Casey}, {Crawford},
  {Dencheva}, {Ely}, {Jenness}, {Labrie}, {Lim}, {Pierfederici}, {Pontzen},
  {Ptak}, {Refsdal}, {Servillat}, \& {Streicher}}]{astropy:2013}
{Astropy Collaboration}, {Robitaille}, T.~P., {Tollerud}, E.~J., {et~al.} 2013,
  \aap, 558, A33, \dodoi{10.1051/0004-6361/201322068}

\bibitem[{{Astropy Collaboration} {et~al.}(2018){Astropy Collaboration},
  {Price-Whelan}, {Sip{H{o}}cz}, {G{"u}nther}, {Lim}, {Crawford}, {Conseil},
  {Shupe}, {Craig}, {Dencheva}, {Ginsburg}, {Vand erPlas}, {Bradley},
  {P{'e}rez-Su{'a}rez}, {de Val-Borro}, {Aldcroft}, {Cruz}, {Robitaille},
  {Tollerud}, {Ardelean}, {Babej}, {Bach}, {Bachetti}, {Bakanov}, {Bamford},
  {Barentsen}, {Barmby}, {Baumbach}, {Berry}, {Biscani}, {Boquien}, {Bostroem},
  {Bouma}, {Brammer}, {Bray}, {Breytenbach}, {Buddelmeijer}, {Burke},
  {Calderone}, {Cano Rodr{'i}guez}, {Cara}, {Cardoso}, {Cheedella}, {Copin},
  {Corrales}, {Crichton}, {D'Avella}, {Deil}, {Depagne}, {Dietrich}, {Donath},
  {Droettboom}, {Earl}, {Erben}, {Fabbro}, {Ferreira}, {Finethy}, {Fox},
  {Garrison}, {Gibbons}, {Goldstein}, {Gommers}, {Greco}, {Greenfield},
  {Groener}, {Grollier}, {Hagen}, {Hirst}, {Homeier}, {Horton}, {Hosseinzadeh},
  {Hu}, {Hunkeler}, {Ivezi{'c}}, {Jain}, {Jenness}, {Kanarek}, {Kendrew},
  {Kern}, {Kerzendorf}, {Khvalko}, {King}, {Kirkby}, {Kulkarni}, {Kumar},
  {Lee}, {Lenz}, {Littlefair}, {Ma}, {Macleod}, {Mastropietro}, {McCully},
  {Montagnac}, {Morris}, {Mueller}, {Mumford}, {Muna}, {Murphy}, {Nelson},
  {Nguyen}, {Ninan}, {N{"o}the}, {Ogaz}, {Oh}, {Parejko}, {Parley}, {Pascual},
  {Patil}, {Patil}, {Plunkett}, {Prochaska}, {Rastogi}, {Reddy Janga},
  {Sabater}, {Sakurikar}, {Seifert}, {Sherbert}, {Sherwood-Taylor}, {Shih},
  {Sick}, {Silbiger}, {Singanamalla}, {Singer}, {Sladen}, {Sooley},
  {Sornarajah}, {Streicher}, {Teuben}, {Thomas}, {Tremblay}, {Turner},
  {Terr{'o}n}, {van Kerkwijk}, {de la Vega}, {Watkins}, {Weaver}, {Whitmore},
  {Woillez}, {Zabalza}, \& {Astropy Contributors}}]{astropy:2018}
{Astropy Collaboration}, {Price-Whelan}, A.~M., {Sip{H{o}}cz}, B.~M., {et~al.}
  2018, aj, 156, 123, \dodoi{10.3847/1538-3881/aabc4f}

\bibitem[{{Barnard} {et~al.}(2008){Barnard}, {Clark}, \& {Kolb}}]{Barnard+08}
{Barnard}, R., {Clark}, J.~S., \& {Kolb}, U.~C. 2008, \aap, 488, 697,
  \dodoi{10.1051/0004-6361:20077975}

\bibitem[{{Binder} {et~al.}(2015){Binder}, {Gross}, {Williams}, \&
  {Simons}}]{Binder+15}
{Binder}, B., {Gross}, J., {Williams}, B.~F., \& {Simons}, D. 2015, \mnras,
  451, 4471, \dodoi{10.1093/mnras/stv1305}

\bibitem[{{Binder} {et~al.}(2011){Binder}, {Williams}, {Eracleous}, {Garcia},
  {Anderson}, \& {Gaetz}}]{Binder+11}
{Binder}, B., {Williams}, B.~F., {Eracleous}, M., {et~al.} 2011, \apj, 742,
  128, \dodoi{10.1088/0004-637X/742/2/128}

\bibitem[{{Binder} {et~al.}(2012){Binder}, {Williams}, {Eracleous}, {Gaetz},
  {Plucinsky}, {Skillman}, {Dalcanton}, {Anderson}, {Weisz}, \&
  {Kong}}]{Binder+12}
---. 2012, \apj, 758, 15, \dodoi{10.1088/0004-637X/758/1/15}

\bibitem[{{Blondin}(1994)}]{Blondin+94}
{Blondin}, J.~M. 1994, \apj, 435, 756, \dodoi{10.1086/174853}

\bibitem[{{Blondin} {et~al.}(1990){Blondin}, {Kallman}, {Fryxell}, \&
  {Taam}}]{Blondin+90}
{Blondin}, J.~M., {Kallman}, T.~R., {Fryxell}, B.~A., \& {Taam}, R.~E. 1990,
  \apj, 356, 591, \dodoi{10.1086/168865}

\bibitem[{{Blondin} {et~al.}(1991){Blondin}, {Stevens}, \&
  {Kallman}}]{Blondin+91}
{Blondin}, J.~M., {Stevens}, I.~R., \& {Kallman}, T.~R. 1991, \apj, 371, 684,
  \dodoi{10.1086/169934}

\bibitem[{{Bondi} \& {Hoyle}(1944)}]{Bondi+44}
{Bondi}, H., \& {Hoyle}, F. 1944, \mnras, 104, 273,
  \dodoi{10.1093/mnras/104.5.273}

\bibitem[{{Carpano} {et~al.}(2019){Carpano}, {Haberl}, {Crowther}, \&
  {Pollock}}]{Carpano+19}
{Carpano}, S., {Haberl}, F., {Crowther}, P., \& {Pollock}, A. 2019, IAU
  Symposium, 346, 187, \dodoi{10.1017/S1743921318007615}

\bibitem[{{Carpano} {et~al.}(2018){Carpano}, {Haberl}, {Maitra}, \&
  {Vasilopoulos}}]{Carpano+18}
{Carpano}, S., {Haberl}, F., {Maitra}, C., \& {Vasilopoulos}, G. 2018, \mnras,
  476, L45, \dodoi{10.1093/mnrasl/sly030}

\bibitem[{{Carpano} {et~al.}(2007){Carpano}, {Pollock}, {Prestwich},
  {Crowther}, {Wilms}, {Yungelson}, \& {Ehle}}]{Carpano+07}
{Carpano}, S., {Pollock}, A.~M.~T., {Prestwich}, A., {et~al.} 2007, \aap, 466,
  L17, \dodoi{10.1051/0004-6361:20077363}

\bibitem[{{Castor} {et~al.}(1975){Castor}, {Abbott}, \& {Klein}}]{Castor+75}
{Castor}, J.~I., {Abbott}, D.~C., \& {Klein}, R.~I. 1975, \apj, 195, 157,
  \dodoi{10.1086/153315}

\bibitem[{{Conti} \& {Morris}(1990)}]{Conti+90}
{Conti}, P.~S., \& {Morris}, P.~W. 1990, \aj, 99, 898, \dodoi{10.1086/115382}

\bibitem[{{Cowley} \& {Crampton}(1975)}]{Cowley+75}
{Cowley}, A.~P., \& {Crampton}, D. 1975, \apjl, 201, L65,
  \dodoi{10.1086/181943}

\bibitem[{{Crowther}(2007)}]{Crowther07}
{Crowther}, P.~A. 2007, \araa, 45, 177,
  \dodoi{10.1146/annurev.astro.45.051806.110615}

\bibitem[{{Crowther} {et~al.}(2010{\natexlab{a}}){Crowther}, {Barnard},
  {Carpano}, {Clark}, {Dhillon}, \& {Pollock}}]{Crowther+10}
{Crowther}, P.~A., {Barnard}, R., {Carpano}, S., {et~al.} 2010{\natexlab{a}},
  \mnras, 403, L41, \dodoi{10.1111/j.1745-3933.2010.00811.x}

\bibitem[{{Crowther} {et~al.}(2007){Crowther}, {Carpano}, {Hadfield}, \&
  {Pollock}}]{Crowther+07}
{Crowther}, P.~A., {Carpano}, S., {Hadfield}, L.~J., \& {Pollock}, A.~M.~T.
  2007, \aap, 469, L31, \dodoi{10.1051/0004-6361:20077677}

\bibitem[{{Crowther} {et~al.}(2010{\natexlab{b}}){Crowther}, {Schnurr},
  {Hirschi}, {Yusof}, {Parker}, {Goodwin}, \& {Kassim}}]{Crowther+10b}
{Crowther}, P.~A., {Schnurr}, O., {Hirschi}, R., {et~al.} 2010{\natexlab{b}},
  \mnras, 408, 731, \dodoi{10.1111/j.1365-2966.2010.17167.x}

\bibitem[{{Dalcanton} {et~al.}(2009){Dalcanton}, {Williams}, {Seth}, {Dolphin},
  {Holtzman}, {Rosema}, {Skillman}, {Cole}, {Girardi}, {Gogarten},
  {Karachentsev}, {Olsen}, {Weisz}, {Christensen}, {Freeman}, {Gilbert},
  {Gallart}, {Harris}, {Hodge}, {de Jong}, {Karachentseva}, {Mateo}, {Stetson},
  {Tavarez}, {Zaritsky}, {Governato}, \& {Quinn}}]{Dalcanton+09}
{Dalcanton}, J.~J., {Williams}, B.~F., {Seth}, A.~C., {et~al.} 2009, \apjs,
  183, 67, \dodoi{10.1088/0067-0049/183/1/67}

\bibitem[{{Day} \& {Stevens}(1993)}]{Day+93}
{Day}, C.~S.~R., \& {Stevens}, I.~R. 1993, \apj, 403, 322,
  \dodoi{10.1086/172205}

\bibitem[{{de Jong} {et~al.}(1996){de Jong}, {van Paradijs}, \&
  {Augusteijn}}]{deJong+96}
{de Jong}, J.~A., {van Paradijs}, J., \& {Augusteijn}, T. 1996, \aap, 314, 484

\bibitem[{{Doran} {et~al.}(2013){Doran}, {Crowther}, {de Koter}, {Evans},
  {McEvoy}, {Walborn}, {Bastian}, {Bestenlehner}, {Gr{\"a}fener}, {Herrero},
  {K{\"o}hler}, {Ma{\'\i}z Apell{\'a}niz}, {Najarro}, {Puls}, {Sana},
  {Schneider}, {Taylor}, {van Loon}, \& {Vink}}]{Doran+13}
{Doran}, E.~I., {Crowther}, P.~A., {de Koter}, A., {et~al.} 2013, \aap, 558,
  A134, \dodoi{10.1051/0004-6361/201321824}

\bibitem[{{Earnshaw} \& {Roberts}(2017)}]{Earnshaw+17}
{Earnshaw}, H.~M., \& {Roberts}, T.~P. 2017, \mnras, 467, 2690,
  \dodoi{10.1093/mnras/stx308}

\bibitem[{{Edgar}(2004)}]{Edgar+04}
{Edgar}, R. 2004, \nar, 48, 843, \dodoi{10.1016/j.newar.2004.06.001}

\bibitem[{{Eggleton}(1983)}]{Eggleton+83}
{Eggleton}, P.~P. 1983, \apj, 268, 368, \dodoi{10.1086/160960}

\bibitem[{{El Mellah} {et~al.}(2019{\natexlab{a}}){El Mellah}, {Sander},
  {Sundqvist}, \& {Keppens}}]{ElMellah+19a}
{El Mellah}, I., {Sander}, A.~A.~C., {Sundqvist}, J.~O., \& {Keppens}, R.
  2019{\natexlab{a}}, \aap, 622, A189, \dodoi{10.1051/0004-6361/201834498}

\bibitem[{{El Mellah} {et~al.}(2019{\natexlab{b}}){El Mellah}, {Sundqvist}, \&
  {Keppens}}]{ElMellah+19b}
{El Mellah}, I., {Sundqvist}, J.~O., \& {Keppens}, R. 2019{\natexlab{b}}, \aap,
  622, L3, \dodoi{10.1051/0004-6361/201834543}

\bibitem[{{Fragos} \& {McClintock}(2015)}]{Fragos+15}
{Fragos}, T., \& {McClintock}, J.~E. 2015, \apj, 800, 17,
  \dodoi{10.1088/0004-637X/800/1/17}

\bibitem[{{Fransson} \& {Fabian}(1980)}]{Fransson+80}
{Fransson}, C., \& {Fabian}, A.~C. 1980, \aap, 87, 102

\bibitem[{{Fruscione} {et~al.}(2006){Fruscione}, {McDowell}, {Allen},
  {Brickhouse}, {Burke}, {Davis}, {Durham}, {Elvis}, {Galle}, {Harris},
  {Huenemoerder}, {Houck}, {Ishibashi}, {Karovska}, {Nicastro}, {Noble},
  {Nowak}, {Primini}, {Siemiginowska}, {Smith}, \& {Wise}}]{Fruscione+06}
{Fruscione}, A., {McDowell}, J.~C., {Allen}, G.~E., {et~al.} 2006, in Society
  of Photo-Optical Instrumentation Engineers (SPIE) Conference Series, Vol.
  6270, Society of Photo-Optical Instrumentation Engineers (SPIE) Conference
  Series, ed. D.~R. {Silva} \& R.~E. {Doxsey}, 62701V,
  \dodoi{10.1117/12.671760}

\bibitem[{{Gehrels}(1986)}]{Gehrels86}
{Gehrels}, N. 1986, \apj, 303, 336, \dodoi{10.1086/164079}

\bibitem[{{Gierli{\'n}ski} {et~al.}(2009){Gierli{\'n}ski}, {Done}, \&
  {Page}}]{Gierlinski+09}
{Gierli{\'n}ski}, M., {Done}, C., \& {Page}, K. 2009, \mnras, 392, 1106,
  \dodoi{10.1111/j.1365-2966.2008.14166.x}

\bibitem[{{Gim{\'e}nez-Garc{\'\i}a} {et~al.}(2016){Gim{\'e}nez-Garc{\'\i}a},
  {Shenar}, {Torrej{\'o}n}, {Oskinova}, {Mart{\'\i}nez-N{\'u}{\~n}ez},
  {Hamann}, {Rodes-Roca}, {Gonz{\'a}lez-Gal{\'a}n}, {Alonso-Santiago},
  {Gonz{\'a}lez-Fern{\'a}ndez}, {Bernabeu}, \& {Sander}}]{Gimenez-Garcia+16}
{Gim{\'e}nez-Garc{\'\i}a}, A., {Shenar}, T., {Torrej{\'o}n}, J.~M., {et~al.}
  2016, \aap, 591, A26, \dodoi{10.1051/0004-6361/201527551}

\bibitem[{{Ginsburg} \& {Mirocha}(2011)}]{Ginsburg+11}
{Ginsburg}, A., \& {Mirocha}, J. 2011, {PySpecKit: Python Spectroscopic
  Toolkit}.
\newblock \doeprint{1109.001}

\bibitem[{{Hamann} \& {Gr{\"a}fener}(2004)}]{Hamann+04}
{Hamann}, W.~R., \& {Gr{\"a}fener}, G. 2004, \aap, 427, 697,
  \dodoi{10.1051/0004-6361:20040506}

\bibitem[{{Hamann} {et~al.}(2006){Hamann}, {Gr{\"a}fener}, \&
  {Liermann}}]{Hamann+06}
{Hamann}, W.~R., {Gr{\"a}fener}, G., \& {Liermann}, A. 2006, \aap, 457, 1015,
  \dodoi{10.1051/0004-6361:20065052}

\bibitem[{{Hamann} \& {Koesterke}(1998)}]{Hamann+98}
{Hamann}, W.~R., \& {Koesterke}, L. 1998, \aap, 335, 1003

\bibitem[{{Hatchett} \& {McCray}(1977)}]{Hatchett+77}
{Hatchett}, S., \& {McCray}, R. 1977, \apj, 211, 552, \dodoi{10.1086/154962}

\bibitem[{{HI4PI Collaboration} {et~al.}(2016){HI4PI Collaboration}, {Ben
  Bekhti}, {Fl{\"o}er}, {Keller}, {Kerp}, {Lenz}, {Winkel}, {Bailin},
  {Calabretta}, {Dedes}, {Ford}, {Gibson}, {Haud}, {Janowiecki}, {Kalberla},
  {Lockman}, {McClure-Griffiths}, {Murphy}, {Nakanishi}, {Pisano}, \&
  {Staveley-Smith}}]{HI4PICollaboration+16}
{HI4PI Collaboration}, {Ben Bekhti}, N., {Fl{\"o}er}, L., {et~al.} 2016, \aap,
  594, A116, \dodoi{10.1051/0004-6361/201629178}

\bibitem[{Hoard {et~al.}(2010)Hoard, Lu, Knigge, Homer, Szkody, Still, Long,
  Dhillon, \& Wachter}]{Hoard+10}
Hoard, D.~W., Lu, T.-N., Knigge, C., {et~al.} 2010, The Astronomical Journal,
  140, 1313, \dodoi{10.1088/0004-6256/140/5/1313}

\bibitem[{{Huarte-Espinosa} {et~al.}(2013){Huarte-Espinosa},
  {Carroll-Nellenback}, {Nordhaus}, {Frank}, \&
  {Blackman}}]{Huarte-Espinosa+13}
{Huarte-Espinosa}, M., {Carroll-Nellenback}, J., {Nordhaus}, J., {Frank}, A.,
  \& {Blackman}, E.~G. 2013, \mnras, 433, 295, \dodoi{10.1093/mnras/stt725}

\bibitem[{{Kaper} {et~al.}(1994){Kaper}, {Hammerschlag-Hensberge}, \&
  {Zuiderwijk}}]{Kaper+94}
{Kaper}, L., {Hammerschlag-Hensberge}, G., \& {Zuiderwijk}, E.~J. 1994, \aap,
  289, 846

\bibitem[{{Koliopanos} {et~al.}(2019){Koliopanos}, {Vasilopoulos}, {Buchner},
  {Maitra}, \& {Haberl}}]{Koliopanos+19}
{Koliopanos}, F., {Vasilopoulos}, G., {Buchner}, J., {Maitra}, C., \& {Haberl},
  F. 2019, \aap, 621, A118, \dodoi{10.1051/0004-6361/201834144}

\bibitem[{{Koljonen} \& {Maccarone}(2017)}]{Koljonen+17}
{Koljonen}, K.~I.~I., \& {Maccarone}, T.~J. 2017, \mnras, 472, 2181,
  \dodoi{10.1093/mnras/stx2106}

\bibitem[{{Krti{\v{c}}ka} {et~al.}(2015){Krti{\v{c}}ka}, {Kub{\'a}t}, \&
  {Krti{\v{c}}kov{\'a}}}]{Krticka+15}
{Krti{\v{c}}ka}, J., {Kub{\'a}t}, J., \& {Krti{\v{c}}kov{\'a}}, I. 2015, \aap,
  579, A111, \dodoi{10.1051/0004-6361/201525637}

\bibitem[{{Kurosawa} {et~al.}(2002){Kurosawa}, {Hillier}, \&
  {Pittard}}]{Kurosawa+02}
{Kurosawa}, R., {Hillier}, D.~J., \& {Pittard}, J.~M. 2002, \aap, 388, 957,
  \dodoi{10.1051/0004-6361:20020443}

\bibitem[{{Lauberts} \& {Valentijn}(1989)}]{Lauberts+89}
{Lauberts}, A., \& {Valentijn}, E.~A. 1989, {The surface photometry catalogue
  of the ESO-Uppsala galaxies}

\bibitem[{{Laycock} {et~al.}(2015{\natexlab{a}}){Laycock}, {Cappallo}, \&
  {Moro}}]{Laycock+15a}
{Laycock}, S. G.~T., {Cappallo}, R.~C., \& {Moro}, M.~J. 2015{\natexlab{a}},
  \mnras, 446, 1399, \dodoi{10.1093/mnras/stu2151}

\bibitem[{{Laycock} {et~al.}(2015{\natexlab{b}}){Laycock}, {Maccarone}, \&
  {Christodoulou}}]{Laycock+15b}
{Laycock}, S. G.~T., {Maccarone}, T.~J., \& {Christodoulou}, D.~M.
  2015{\natexlab{b}}, \mnras, 452, L31, \dodoi{10.1093/mnrasl/slv082}

\bibitem[{{Leitherer} {et~al.}(2019){Leitherer}, {Lee}, \&
  {Faisst}}]{Leitherer+19}
{Leitherer}, C., {Lee}, J.~C., \& {Faisst}, A. 2019, \aj, 158, 192,
  \dodoi{10.3847/1538-3881/ab44c0}

\bibitem[{{L{\'e}pine} {et~al.}(2000){L{\'e}pine}, {Moffat}, {St-Louis},
  {Marchenko}, {Dalton}, {Crowther}, {Smith}, {Willis}, {Antokhin}, \&
  {Tovmassian}}]{Lepine+00}
{L{\'e}pine}, S., {Moffat}, A. F.~J., {St-Louis}, N., {et~al.} 2000, \aj, 120,
  3201, \dodoi{10.1086/316858}

\bibitem[{{Lucy} \& {Solomon}(1970)}]{Lucy+70}
{Lucy}, L.~B., \& {Solomon}, P.~M. 1970, \apj, 159, 879, \dodoi{10.1086/150365}

\bibitem[{{Madura} {et~al.}(2013){Madura}, {Gull}, {Okazaki}, {Russell},
  {Owocki}, {Groh}, {Corcoran}, {Hamaguchi}, \& {Teodoro}}]{Madura+13}
{Madura}, T.~I., {Gull}, T.~R., {Okazaki}, A.~T., {et~al.} 2013, \mnras, 436,
  3820, \dodoi{10.1093/mnras/stt1871}

\bibitem[{{Manousakis} \& {Walter}(2015)}]{Manousakis+15}
{Manousakis}, A., \& {Walter}, R. 2015, \aap, 584, A25,
  \dodoi{10.1051/0004-6361/201526893}

\bibitem[{{Marchenko} {et~al.}(2004){Marchenko}, {Moffat}, {Crowther},
  {Chen{\'e}}, {De Serres}, {Eenens}, {Hill}, {Moran}, \&
  {Morel}}]{Marchenko+04}
{Marchenko}, S.~V., {Moffat}, A.~F.~J., {Crowther}, P.~A., {et~al.} 2004,
  \mnras, 353, 153, \dodoi{10.1111/j.1365-2966.2004.08058.x}

\bibitem[{{Marsh} \& {Horne}(1990)}]{Marsh+90a}
{Marsh}, T.~R., \& {Horne}, K. 1990, \apj, 349, 593, \dodoi{10.1086/168346}

\bibitem[{{Marsh} {et~al.}(1990){Marsh}, {Horne}, {Schlegel}, {Honeycutt}, \&
  {Kaitchuck}}]{Marsh+90b}
{Marsh}, T.~R., {Horne}, K., {Schlegel}, E.~M., {Honeycutt}, R.~K., \&
  {Kaitchuck}, R.~H. 1990, \apj, 364, 637, \dodoi{10.1086/169446}

\bibitem[{{Martin} {et~al.}(2006){Martin}, {Davidson}, {Humphreys}, {Hillier},
  \& {Ishibashi}}]{Martin+06}
{Martin}, J.~C., {Davidson}, K., {Humphreys}, R.~M., {Hillier}, D.~J., \&
  {Ishibashi}, K. 2006, \apj, 640, 474, \dodoi{10.1086/500038}

\bibitem[{{Mart{\'\i}nez-N{\'u}{\~n}ez}
  {et~al.}(2017){Mart{\'\i}nez-N{\'u}{\~n}ez}, {Kretschmar}, {Bozzo},
  {Oskinova}, {Puls}, {Sidoli}, {Sundqvist}, {Blay}, {Falanga}, {F{\"u}rst},
  {G{\'\i}menez-Garc{\'\i}a}, {Kreykenbohm}, {K{\"u}hnel}, {Sander},
  {Torrej{\'o}n}, \& {Wilms}}]{MartinezNunez+17}
{Mart{\'\i}nez-N{\'u}{\~n}ez}, S., {Kretschmar}, P., {Bozzo}, E., {et~al.}
  2017, \ssr, 212, 59, \dodoi{10.1007/s11214-017-0340-1}

\bibitem[{{McClintock} {et~al.}(2014){McClintock}, {Narayan}, \&
  {Steiner}}]{McClintock+14}
{McClintock}, J.~E., {Narayan}, R., \& {Steiner}, J.~F. 2014, \ssr, 183, 295,
  \dodoi{10.1007/s11214-013-0003-9}

\bibitem[{{Mehner} {et~al.}(2011){Mehner}, {Davidson}, {Martin}, {Humphreys},
  {Ishibashi}, \& {Ferland }}]{Mehner+11}
{Mehner}, A., {Davidson}, K., {Martin}, J.~C., {et~al.} 2011, \apj, 740, 80,
  \dodoi{10.1088/0004-637X/740/2/80}

\bibitem[{{Moffat} {et~al.}(1988){Moffat}, {Drissen}, {Lamontagne}, \&
  {Robert}}]{Moffat+88}
{Moffat}, A. F.~J., {Drissen}, L., {Lamontagne}, R., \& {Robert}, C. 1988,
  \apj, 334, 1038, \dodoi{10.1086/166895}

\bibitem[{{Montgomery}(2012)}]{Montgomery+12}
{Montgomery}, M.~M. 2012, \apjl, 745, L25, \dodoi{10.1088/2041-8205/745/2/L25}

\bibitem[{{{\"O}zel} {et~al.}(2010){{\"O}zel}, {Psaltis}, {Narayan}, \&
  {McClintock}}]{Ozel+10}
{{\"O}zel}, F., {Psaltis}, D., {Narayan}, R., \& {McClintock}, J.~E. 2010,
  \apj, 725, 1918, \dodoi{10.1088/0004-637X/725/2/1918}

\bibitem[{{Podsiadlowski} \& {Mohamed}(2007)}]{Podsiadlowski+07}
{Podsiadlowski}, P., \& {Mohamed}, S. 2007, Baltic Astronomy, 16, 26

\bibitem[{{Prestwich} {et~al.}(2007){Prestwich}, {Kilgard}, {Crowther},
  {Carpano}, {Pollock}, {Zezas}, {Saar}, {Roberts}, \& {Ward}}]{Prestwich+07}
{Prestwich}, A.~H., {Kilgard}, R., {Crowther}, P.~A., {et~al.} 2007, \apjl,
  669, L21, \dodoi{10.1086/523755}

\bibitem[{{Rutten} {et~al.}(1992){Rutten}, {van Paradijs}, \&
  {Tinbergen}}]{Rutten+92}
{Rutten}, R.~G.~M., {van Paradijs}, J., \& {Tinbergen}, J. 1992, \aap, 260, 213

\bibitem[{{Schmutz} {et~al.}(1989){Schmutz}, {Hamann}, \&
  {Wessolowski}}]{Schmutz+89}
{Schmutz}, W., {Hamann}, W.~R., \& {Wessolowski}, U. 1989, \aap, 210, 236

\bibitem[{{Schulz} {et~al.}(2002){Schulz}, {Canizares}, {Lee}, \&
  {Sako}}]{Schulz+02}
{Schulz}, N.~S., {Canizares}, C.~R., {Lee}, J.~C., \& {Sako}, M. 2002, \apjl,
  564, L21, \dodoi{10.1086/338862}

\bibitem[{{Silverman} \& {Filippenko}(2008)}]{Silverman+08}
{Silverman}, J.~M., \& {Filippenko}, A.~V. 2008, \apjl, 678, L17,
  \dodoi{10.1086/588096}

\bibitem[{{St.-Louis} {et~al.}(1993){St.-Louis}, {Howarth}, {Willis},
  {Stickland }, {Smith}, {Conti}, \& {Garmany}}]{St-Louis+93}
{St.-Louis}, N., {Howarth}, I.~D., {Willis}, A.~J., {et~al.} 1993, \aap, 267,
  447

\bibitem[{{Stanishev} {et~al.}(2004){Stanishev}, {Kraicheva}, {Boffin},
  {Genkov}, {Papadaki}, \& {Carpano}}]{Stanishev+04}
{Stanishev}, V., {Kraicheva}, Z., {Boffin}, H.~M.~J., {et~al.} 2004, \aap, 416,
  1057, \dodoi{10.1051/0004-6361:20034145}

\bibitem[{{Steeghs} \& {Casares}(2002)}]{Steeghs+02}
{Steeghs}, D., \& {Casares}, J. 2002, \apj, 568, 273, \dodoi{10.1086/339224}

\bibitem[{{Steiner} {et~al.}(2014){Steiner}, {McClintock}, {Orosz},
  {Remillard}, {Bailyn}, {Kolehmainen}, \& {Straub}}]{Steiner+14}
{Steiner}, J.~F., {McClintock}, J.~E., {Orosz}, J.~A., {et~al.} 2014, \apjl,
  793, L29, \dodoi{10.1088/2041-8205/793/2/L29}

\bibitem[{{Suleimanov} {et~al.}(2003){Suleimanov}, {Meyer}, \&
  {Meyer-Hofmeister}}]{Suleimanov+03}
{Suleimanov}, V., {Meyer}, F., \& {Meyer-Hofmeister}, E. 2003, \aap, 401, 1009,
  \dodoi{10.1051/0004-6361:20030159}

\bibitem[{{Teodoro} {et~al.}(2012){Teodoro}, {Damineli}, {Arias}, {de
  Ara{\'u}jo}, {Barb{\'a}}, {Corcoran}, {Borges Fernandes},
  {Fern{\'a}ndez-Laj{\'u}s}, {Fraga}, {Gamen}, {Gonz{\'a}lez}, {Groh},
  {Marshall}, {McGregor}, {Morrell}, {Nicholls}, {Parkin}, {Pereira},
  {Phillips}, {Solivella}, {Steiner}, {Stritzinger}, {Thompson}, {Torres},
  {Torres}, \& {Zevallos Herencia}}]{Teodoro+12}
{Teodoro}, M., {Damineli}, A., {Arias}, J.~I., {et~al.} 2012, \apj, 746, 73,
  \dodoi{10.1088/0004-637X/746/1/73}

\bibitem[{{Teodoro} {et~al.}(2016){Teodoro}, {Damineli}, {Heathcote},
  {Richardson}, {Moffat}, {St-Jean}, {Russell}, {Gull}, {Madura}, {Pollard},
  {Walter}, {Coimbra}, {Prates}, {Fern{\'a}ndez-Laj{\'u}s}, {Gamen}, {Hickel},
  {Henrique}, {Navarete}, {Andrade}, {Jablonski}, {Luckas}, {Locke}, {Powles},
  {Bohlsen}, {Chini}, {Corcoran}, {Hamaguchi}, {Groh}, {Hillier}, \&
  {Weigelt}}]{Teodoro+16}
{Teodoro}, M., {Damineli}, A., {Heathcote}, B., {et~al.} 2016, \apj, 819, 131,
  \dodoi{10.3847/0004-637X/819/2/131}

\bibitem[{{Todt} {et~al.}(2015){Todt}, {Sander}, {Hainich}, {Hamann}, {Quade},
  \& {Shenar}}]{Todt+15}
{Todt}, H., {Sander}, A., {Hainich}, R., {et~al.} 2015, \aap, 579, A75,
  \dodoi{10.1051/0004-6361/201526253}

\bibitem[{Tutukov \& Fedorova(2016)}]{Tutukov+16}
Tutukov, A.~V., \& Fedorova, A.~V. 2016, Astronomy Reports, 60, 106,
  \dodoi{10.1134/S1063772915120070}

\bibitem[{{Urbaneja} {et~al.}(2005){Urbaneja}, {Herrero}, {Bresolin},
  {Kudritzki}, {Gieren}, {Puls}, {Przybilla}, {Najarro}, \&
  {Pietrzy{\'n}ski}}]{Urbaneja+05}
{Urbaneja}, M.~A., {Herrero}, A., {Bresolin}, F., {et~al.} 2005, \apj, 622,
  862, \dodoi{10.1086/427468}

\bibitem[{{van Kerkwijk}(1993)}]{vanKerkwijk93}
{van Kerkwijk}, M.~H. 1993, \aap, 276, L9

\bibitem[{{Walton} {et~al.}(2018){Walton}, {Bachetti}, {F{\"u}rst}, {Barret},
  {Brightman}, {Fabian}, {Grefenstette}, {Harrison}, {Heida}, {Kennea},
  {Kosec}, {Lau}, {Madsen}, {Middleton}, {Pinto}, {Steiner}, \&
  {Webb}}]{Walton+18}
{Walton}, D.~J., {Bachetti}, M., {F{\"u}rst}, F., {et~al.} 2018, \apjl, 857,
  L3, \dodoi{10.3847/2041-8213/aabadc}

\bibitem[{{Wolf} \& {Rayet}(1867)}]{Wolf+1867}
{Wolf}, C.~J.~E., \& {Rayet}, G. 1867, Academie des Sciences Paris Comptes
  Rendus, 65, 292

\end{thebibliography}
\bibliographystyle{aasjournal}


\end{document}